# Determination of Semiconductor Diffusion Coefficient by Optical Microscopy Measurements

*Dane W. deQuilettes,[1*] Roberto Brenes,[1,2] Madeleine Laitz,[1,2] Brandon T. Motes,[2] Mikhail M. Glazov,[3] Vladimir Bulović[1,2*]*

[1] Research Laboratory of Electronics, Massachusetts Institute of Technology, 77 Massachusetts Avenue, Cambridge, Massachusetts 02139, USA

[2] Department of Electrical Engineering and Computer Science, Massachusetts Institute of Technology, 77 Massachusetts Avenue, Cambridge, Massachusetts 02139, USA

[3] Ioffe Institute, 26 Politekhnicheskaya, 194021, Saint Petersburg, Russian Federation

*Corresponding Authors: danedeq@mit.edu; bulovic@mit.edu

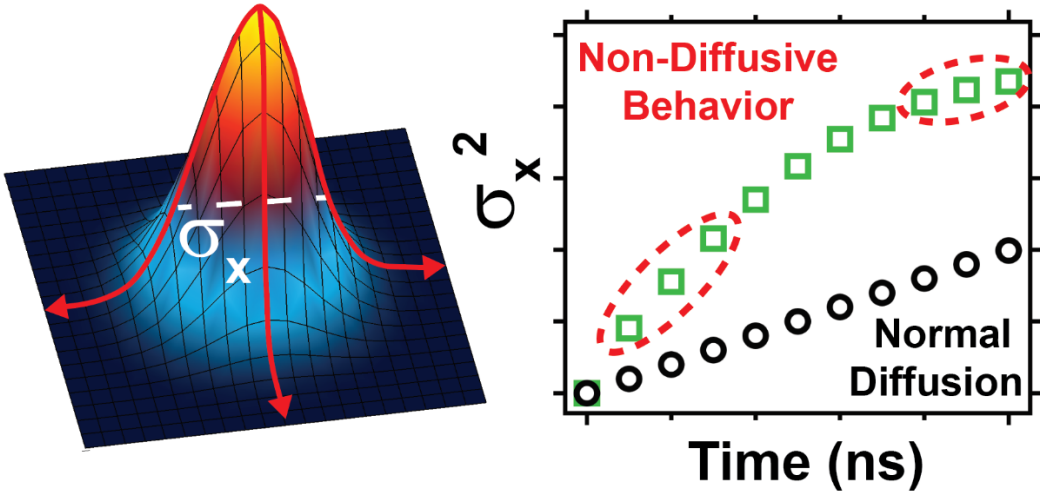

*Energy carrier transport and recombination in emerging semiconductors can be directly monitored with optical microscopy, leading to the measurement of the diffusion coefficient (D), a critical property for design of efficient optoelectronic devices. D is often determined by fitting a time-resolved expanding carrier profile after optical excitation using a Mean Squared Displacement (MSD) Model, where $D = [\sigma^2(t) - \sigma^2(0)]/(2t)$, with $\sigma^2(t)$ being the Gaussian variance as a function of time. Although this approach has gained widespread adoption, its utilization can significantly overestimate D due to the non-linear recombination processes that artificially broaden the carrier distribution profile. Here, we simulate diffusive processes in both excitonic and free carrier semiconductors and present revised MSD Models that take into account second-order (i.e. bimolecular) and third-order (i.e. Auger) processes to accurately recover D for various types of materials. For perovskite thin films, utilization of these models can reduce fitting error by orders of magnitude, especially for commonly deployed excitation conditions where carrier densities are $> 5x10^{16}$ cm$^{-3}$. In addition, we show that commonly-deployed MSD Models are not well-suited for the study of films with microstructure, especially when boundary behavior is unknown and feature sizes are comparable to the diffusion length. Finally, we find that photon recycling only impacts energy carrier profiles on ultrashort time scales or for materials with fast radiative decay times. We present clear strategies to investigate energy transport in disordered materials for more effective design and optimization of electronic and optoelectronic devices.*

**I. INTRODUCTION**

Energy carrier diffusion length and lifetime-mobility product are critical properties of semiconductor materials used for the design of photovoltaics, display technologies, transistors, optical sensors and sources.[1,2] Accurate determination of these values is essential in the fabrication and development of efficient optoelectronic devices. For example, the carrier diffusion length is important in determining the layer and electrode spacing in interdigitated back contact (IBC) device architectures,[3] the target active layer thickness to maximize absorption without sacrificing photogenerated charge collection efficiency,[4] and the bulk heterojunction donor/acceptor ratios for efficient excitonic dissociation at interfaces.[5] As an evaluative optoelectronic parameter, there are now several methods capable of determining diffusion length. Bulk methods for vertical and lateral diffusion include photoluminescence (PL) quenching,[6] terahertz or microwave conductivity,[7,8] Hall effect measurements,[9] and transient grating/diffractive optics experiments.[10]

More recently, the advent of microscopy has allowed for local determination of diffusion rates and lengths, where energetic disorder within the sample such as grain boundaries, and other morphological features can impact local energy transport.[11] The direct visualization of carrier diffusion using both optical and electron microscopies have yielded exciting breakthroughs in semiconductor research. Transient absorption, reflection, and fluorescence microscopies have revealed ballistic transport of hot carriers on short time scales in perovskite thin films.[12-14] Stimulated emission depletion (STED) microscopy has shown exciton migration on its native nanometer length scale in conjugated polymers.[15] Advanced setups such as four-dimensional electron microscopy, where an optical pulse can create femtosecond dynamics for an electron probe, has revealed super-diffusive behavior in doped silicon.[16] All of these methods can offer femtosecond temporal resolution and nanometer spatial resolution, with exciting scope for further advancements. In particular, most optical measurements rely on imaging the expanded energy carrier distributions under a focused (i.e. diffraction-limited) excitation spot and monitoring the expansion over a series of time delays. A kinetic model is then used to fit the diffusion constant and can take into account the diffusion tensor as well as the photoexcited carrier recombination kinetics. This often requires setting up a three-dimensional partial differential equation (PDE) and numerically solving for the excited state density at various time steps through, i.e., finite element analysis. Due to the complexity of this approach, several approximations have been developed to simplify the analysis and reduce the computational cost.

One approach that has gained widespread adoption fits the spread of the photoexcited carrier spatial distribution to Gaussian functions to monitor the change in profile variance (here referred to as the Mean Squared Displacement (MSD) Model).[17] This method has been applied to small molecules,[17,18] silicon,[18] gallium arsenide (GaAs),[19,20] transition metal dichalcogenides (TMDs),[21] and both single crystal and polycrystalline perovskite samples.[22-25] Although this model was first derived for excitonic materials (*i.e.* molecular solids), it is now being used to describe transport processes in many materials with, importantly, widely different recombination mechanisms and kinetics. It has previously been shown that, at high excitation densities, exciton-exciton annihilation can lead to the non-diffusive broadening of the Gaussian profile due to faster recombination at regions of higher exciton density.[26-28] A few recent studies have commented that these higher-order recombination processes can lead to fitting errors in perovskite samples, which may well explain the wide range in reported diffusion coefficient values.[11,29-31] Therefore, approaches that can accurately extract the diffusion coefficient using optical microscopy for emerging semiconductors with different photoexcited species are greatly needed.

Here we apply a theoretical approach informed by material experimental parameters to study carrier broadening effects showing that higher-order processes must be considered in order to accurately extract the diffusion coefficient (*D*) across the commonly reported microscopic experimental conditions. We find that the standard MSD Model currently used is only accurate over a narrow range of excitation fluences and does not capture the impact of sample microstructure on carrier distribution profiles. Even in what many consider the low fluence (i.e. $< 1$ µJ cm$^{-2}$) or low carrier density ($< 1 \times 10^{17}$ cm$^{-3}$) regime, significant errors can still arise in the fitting of the diffusion coefficient due to non-linear recombination processes. In order to overcome these limitations, we derive revised MSD Models taking into account non-linear, second and third-order recombination processes that can accurately recover *D* values measured in the medium and high carrier density regimes. As more studies focus on imaging carrier diffusion over disordered energy landscapes, we expect that this work will enhance the understanding of how to model charge and energy transport mechanisms and accurately derive transport parameters, hence more effectively guiding the design of emerging semiconductors on the microscale.

## II. MEAN SQUARED DISPLACEMENT MODEL FOR FIRST-ORDER RECOMBINATION KINETICS

First, we consider conditions under which the standard MSD Model can be applied. We simulate a common two-dimensional diffusion scenario for a semiconductor that exhibits only first-order kinetics (i.e. radiative and non-radiative excitonic recombination, such as in a molecular solid)

$$\frac{\partial N(\mathbf{u},t)}{\partial t} = \nabla \cdot (D_{ij}(\mathbf{u},t) \nabla N(\mathbf{u},t)) - k_{tot} N(\mathbf{u},t) \tag{1}$$

Here the spatial coordinates are represented by the vector $\mathbf{u} = (x, y)$, $N(\mathbf{u}, t)$ is the spatially and temporally dependent carrier density, $D_{ij}(\mathbf{u}, t)$ is the diffusion coefficient tensor, $\nabla$ is the gradient operator, and $k_{tot}$ is the sum of the radiative ($k_R$) and non-radiative ($k_{NR}$) recombination rate constants (i.e. $k_{tot} = k_R + k_{NR}$). Equation (1) can be numerically solved and the diffusion tensor iteratively fit, but this level of analysis may be excessive and computationally expensive for standard samples. Therefore, for simplicity, the diffusive processes are assumed to be time-independent, isotropic (i.e. $D_{ij} = 0$ when $i \neq j$; $D_{ij} = D$ when $i = j$), and the diffraction-limited laser spot used in confocal microscopy measurements, which resembles an Airy disk, is often approximated as a Gaussian function.[32] These simplifications allow $D$ in Eq. (1) to be analytically solved (see SI for more details)[33]

$$D = \frac{\sigma_x^2(t) - \sigma_x^2(0)}{2t} \quad (2),$$

where the numerator is the change in the mean-squared displacement, $\sigma_x(t)$ is the Gaussian standard deviation of the measured intensity profile along the arbitrary spatial dimension $x$ as a function of time ($t$), and $\sigma_x^2$ is the variance. Oftentimes, we do not have direct access to the carrier distribution profile, but rather measure a spectroscopic signal that is proportional to the distribution. Throughout the rest of the text, we use $\sigma_x^2$ to describe the variance in the simulated and/or measured signal intensity (see SI for more details).

Advantages of Eq. (2) are that it allows $D$ to be extracted by simply fitting the slope ($m$) of the linear MSD curve with respect to time where ($D = m/2$). It can also be used to quickly determine if transport behavior deviates from normal diffusive behavior and it removes any broadening contributions due to the point-spread-function (PSF) of the measurement setup.[34] We note that an $\alpha$ parameter, an exponential modifier of time $t$ [i.e. replace $t$ by $t^\alpha$ in the denominator of Eq. (2)], is often used as a free variable in this equation to quantify energetic disorder or sub-diffusive transport,[34] here we only consider normal diffusion where $\alpha = 1$, hence $\alpha$ is not present in Eq. (2).

Figure 1 shows the result of this model when applied to simulated PL intensity maps, where the PL $\sim k_R N$ (for simplicity we assume $k_R \gg k_{NR}$ and therefore $k_{tot} \sim k_R$). We first use Eq. (1) to generate a 2-dimensional set of PL data with a commonly reported diffusion coefficient of 0.05 cm$^2$/s,[30,35] and refer to it as $D_{sim}$. At each time step of the simulation, we fit the normalized PL profile with a Gaussian function of the form $N(x,t) = Ae^{-x^2/2\sigma_x^2}$ and extract $\sigma_x^2(t)$ as the free variable. We then plot the mean squared displacement $(\sigma_x^2(t) - \sigma_x^2(0))$ as a function of $t$ and fit the data to Eq. (2), which was separately derived, to extract the fitted diffusion coefficient ($D_{fit}$). In other words, we generated a realistic data set with physical parameters that have been reported in the literature,[36] and have *a-priori* knowledge of the $D$ value we seek to recover through our fit of Eq. (2). Figure 1a-c show the PL maps over typical time intervals of 0, 10, and 20 ns with an initial integrated carrier density ($N_0$) of $1\times10^{16}$ cm$^{-3}$,[25,30] where we observe the diffusion of carriers outside of the initial Gaussian excitation spot (which has full width at half maximum (FWHM) of 660 nm) and a decrease in emission intensity due to first-order ($k_{tot}$) recombination (see Figure S1 for similar simulations assuming an Airy disk excitation distribution).[33] Figure 1d shows normalized PL intensity line profiles of the generated data along the *x*-axis (red dashed line in Figure 1a) as a function of time, where the Gaussian standard deviation ($\sigma_x$) is also shown. Figure 1e shows the fit to the mean squared displacement as a function of time using Eq. (2), where $D_{sim} = D_{fit} = 0.05$ cm$^2$s$^{-1}$. This result demonstrates that Eq. (2) can be used to accurately extract the $D$ value input into the simulation and can be applied to samples with isotropic diffusion where the signal (i.e. PL) ~ $k_R N$. Although Eq. (2) appears to accurately yield $D$ for materials which display simple first-order recombination kinetics (as would be found in the excitonic solids, such as molecular films), it is unclear whether this same framework can be used for photogenerated free carriers in a semiconductor, where the signal intensity ~ $kN^2$, and where higher-order recombination processes might be present.

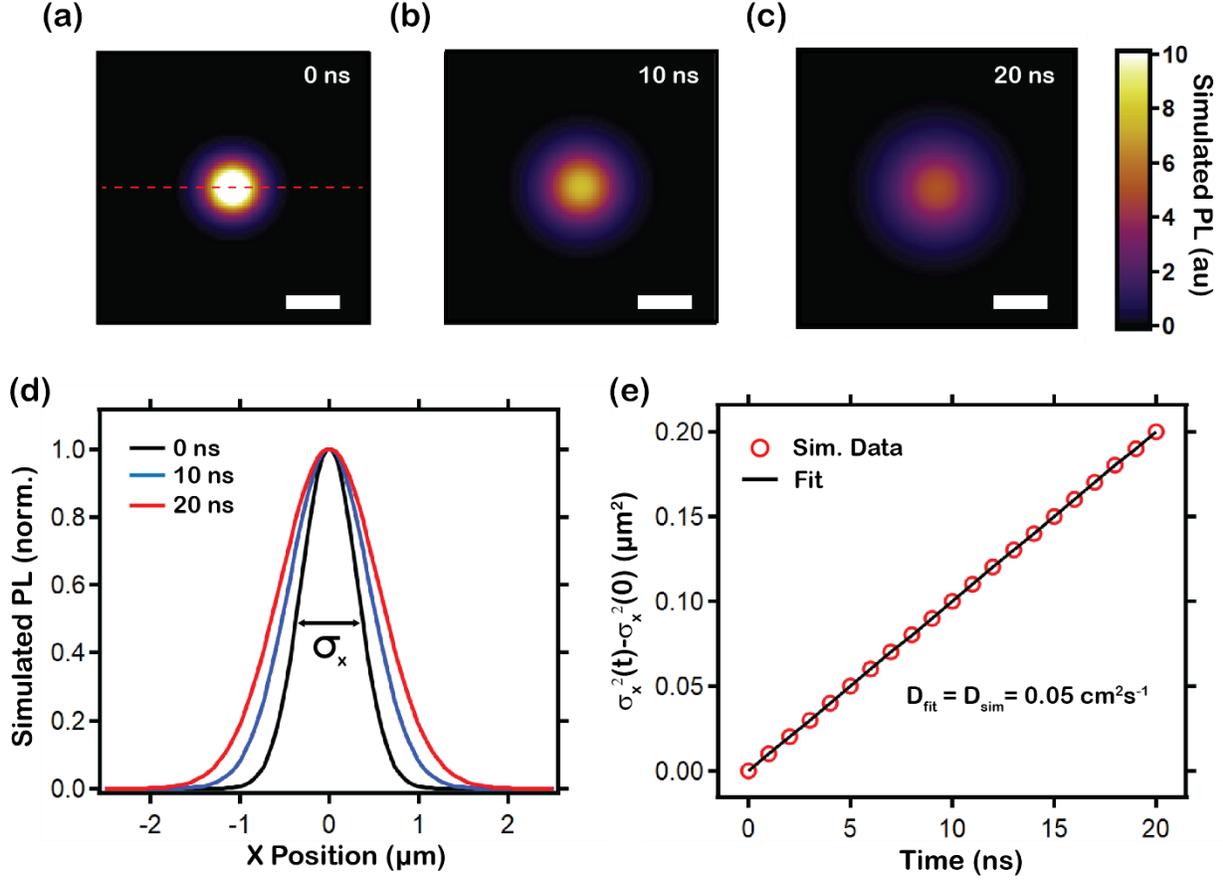

**Figure 1.** a-c) Simulated photoluminescence (PL) diffusion maps following first-order ($k_{tot}N$) recombination kinetics (domain size = 5 x 5 μm; $N_0$ = 1x10$^{16}$ cm$^{-3}$; $k_{tot}$ = 1x10$^6$ s$^{-1}$; $D_{sim}$ = 0.05 cm$^2$ s$^{-1}$; PL ~ $k_R N$, where $k_{tot}$ ~ $k_R$), scale bars are 1 μm. d) Temporal line profiles along the x-direction, indicated by the red dashed line shown in (a). e) Mean squared displacement (MSD) as function of time along with a fit using the MSD Model, where $D_{fit} = [\sigma_x^2(t) - \sigma_x^2(0)]/2t$, Eq. (2). The fitted diffusion coefficient ($D_{fit}$ = 0.05 cm$^2$ s$^{-1}$) matches the pre-defined $D_{sim}$ value used to generate the simulated PL diffusion data.

## III. MSD MODEL FOR SECOND-ORDER RECOMBINATION KINETICS

Equation 2 has been applied to both excitonic and free carrier semiconductors.[11,17,22,23,34] In excitonic materials measured at low carrier densities, the measured signal primarily scales as $N$, until exciton-exciton annihilation effects begin to dominate.[26] Although higher-order effects are sometimes neglected due to the low non-linear recombination coefficients, at higher carrier densities and especially in free carrier semiconductors, these effects become more pronounced and must be considered.[30,37,38] In order to quantify these effects, we use the non-linear diffusion equation to model the impact of second-order recombination on the spatiotemporal carrier density:

$$\frac{\partial N(\boldsymbol{u},t)}{\partial t} = \nabla \cdot (D_{ij}(\boldsymbol{u})\nabla N(\boldsymbol{u},t)) - k_1 N(\boldsymbol{u},t) - k_2 N^2(\boldsymbol{u},t) \qquad (3)$$

where $k_1$ is the non-radiative, first-order (monomolecular) recombination constant and $k_2$ is the second-order (bimolecular) recombination rate constant. Equation (3) can be applied for free carriers and has the same form for excitonic materials, but in the latter case the coefficients $k_1$ and $k_2$ have different definitions.[26] Here, as in the previous section, the carrier diffusivity is assumed to be isotropic and the same for electrons and holes.

Figure 2a shows simulated carrier density maps using Eq. (3) with and without the second-order recombination term ($k_2N^2$) at 0 and 10 ns for a fixed $D$ of 0.05 cm$^2$ s$^{-1}$. In order to emphasize the differences in the two maps at various time slices, Figure 2b shows the normalized spatial profiles where, importantly, the line profiles appear broader for the case with the nonlinear term. These differences become more evident when monitoring the MSD values as a function of time (Fig. 2c), where the simulated data without $k_2N^2$ appears linear and identical to Fig. 1, but shows nonlinear behavior when $k_2N^2$ is included. We note for cases with nonlinear recombination, the term mean-squared-displacement is not entirely accurate because the profile changes cannot be only attributed to overall changes in position.

We highlight that the diffusion coefficient is the exact same in these simulations and the only variable changing is whether second-order recombination is included or not. These results highlight that recombination mechanisms have a large impact on the MSD magnitudes and line shapes. As highlighted by other groups,[26,29] nonlinear recombination leads to a flattening of the carrier profile due to faster recombination at the center of the excitation spot.

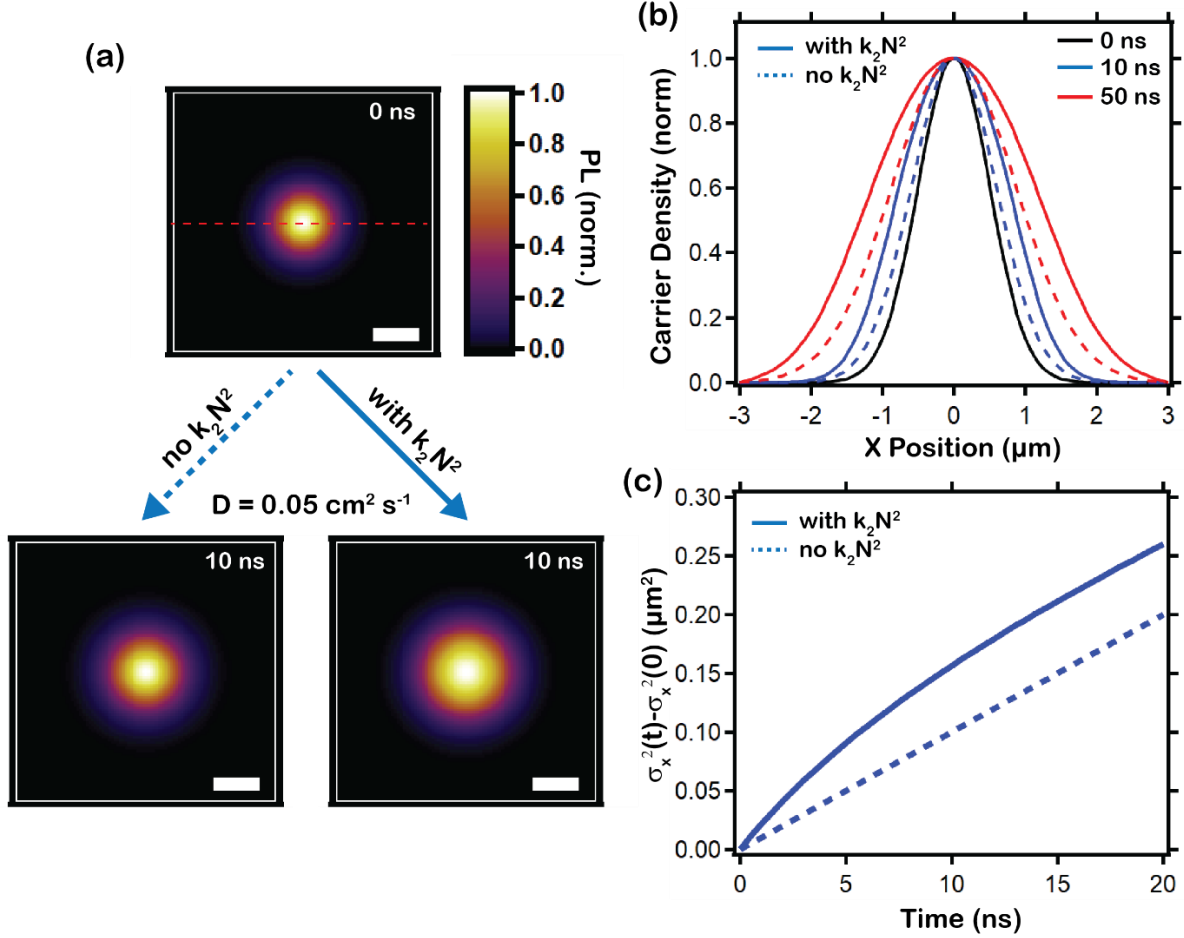

**Figure 2.** a) Simulated carrier density diffusion maps for a fixed diffusion coefficient ($D_{sim}$) and taking into account only first-order (no $k_2N^2$) as well as first and second order (with $k_2N^2$) recombination kinetics. Domain size = 6 x 6 $\mu$m, $N_0 = 1\times10^{18}$ cm$^{-3}$, $k_{tot} = 1\times10^6$ s$^{-1}$, $k_2 = 2\times10^{-10}$ cm$^3$ s$^{-1}$, $D_{sim} = 0.05$ cm$^2$ s$^{-1}$, and scale bars are 1 $\mu$m. b) Temporal line profiles along the x-direction, indicated by the red dashed line shown in (a). c) Mean squared displacement (MSD) of the profiles in b) as function of time with and without the $k_2N^2$ term.

Figure 2 shows that at higher carrier densities, non-linear processes can dominate the shape of the MSD profile and therefore should be considered when modeling and extracting the diffusion coefficient. In this regard, we follow a similar derivation of the standard MSD model from Eq. (1), and derive a revised MSD Model shown in Eq. (4), which accounts for second-order recombination with some approximations (see SI for more details)[33].

$$D = \frac{\sigma_{x,2}^2(t) - \sigma_{x,2}^2(0) - \int_0^t \frac{k_2 N_{tot}(t')}{16\pi z} dt'}{t} \qquad (4)$$

Here, $N_{tot}(t)$ is the total time-dependent carrier population function with an initial value of $N_{tot}(0) = N_0 \pi \sigma_x^2(0) z$ and

$$N_{tot}(t) = \frac{N_{tot}(0)\exp(-k_1 t)}{1 + \frac{k_2 N_{tot}(0)}{8\pi D z} \exp\left(\frac{\sigma_{x,2}^2(0) k_1}{4D}\right) \left[\text{Ei}\left(-\frac{\sigma_{x,2}^2(0) k_1}{4D} - k_1 t\right) - \text{Ei}\left(-\frac{\sigma_{x,2}^2(0) k_1}{4D}\right)\right]} \qquad (5),$$

where Ei(x) is the exponential integral and $z$ is the film thickness. We use $\sigma_{x,2}^2$ in Eq. (4) to denote the variance being quadratically proportional to the carrier density (see Eqs. S36 in the SI). The last term in the numerator of Eq. (4) accounts for the bimolecular recombination effect. The difference in factor 2 in the denominator as compared with Eq. (2) is related to the definition of the MSD for the radiative bimolecular recombination (cf. Eqs. S33 and S38). For the case of excitons, Eq. (4) holds with $\sigma_{x,2}^2 \to \sigma_x^2$ in the numerator and $t \to 2t$ in the denominator.[26]

Figure 3a shows the fits using Eqs. (4) and (5) to simulated data for initial carrier densities of 1x10[15] cm[-3] and 1x10[18] cm[-3]. At 1x10[15] cm[-3], the behavior is entirely linear $\left(i.e., [\sigma_{x,2}^2(t) - \sigma_{x,2}^2(0)] \gg \int_0^t \frac{k_2 N(t')}{16\pi z} dt'\right)$ and simplifies to a slightly modified version (denominator is "$t$" instead of "$2t$") of Eq. (2) as shown in Eq. (6).

$$D = \frac{\sigma_{x,2}^2(t) - \sigma_{x,2}^2(0)}{t} \qquad (6)$$

At higher carrier densities, the non-linear second term in Eq. (4) begins to dominate and must be included in order to accurately recover $D$ (see Fig. S2). Figure 3b shows fits to the data through a wide range of initial carrier densities, demonstrating that the non-linear MSD Model (Eq. 4) leads to significantly less error compared to the commonly deployed linear MSD Model in Eq. (2). For example, at a carrier density of 5x10$^{18}$ cm$^{-3}$, Eq. (4) underestimates the true value by just -20% and the linear Eq. (6) introduces +200% error. The small offset in $D_{fit}$ values for the nonlinear equation at higher carrier densities is likely attributed to the imperfect ansatz (of Gaussian distribution of the particles) used in the derivation. Recently, several reports have calculated the instantaneous slopes, $(MSD_{n+1} - MSD_n)/(t_{n+1} - t_n)$, of the MSD curves to quantify changes in the diffusion coefficient as a function of time.[14,39] Although, in our case, the intrinsic $D$ is not changing (i.e. $D_{sim}$ is fixed at 0.05 cm$^2$ s$^{-1}$), the apparent diffusion coefficient, $D_{app.} = (MSD_{n+1} - MSD_n)/(t_{n+1} - t_n)$, changes as an artifact resulting from non-linear broadening of the distribution profile (Figure 3c). For our simulation, the artificial broadening becomes negligible around 500 ns, and fitting beyond this point can accurately recover the true $D$ value. Therefore, larger fitting error can be introduced in the earlier times, where second-order recombination processes dominate due to the higher carrier density.

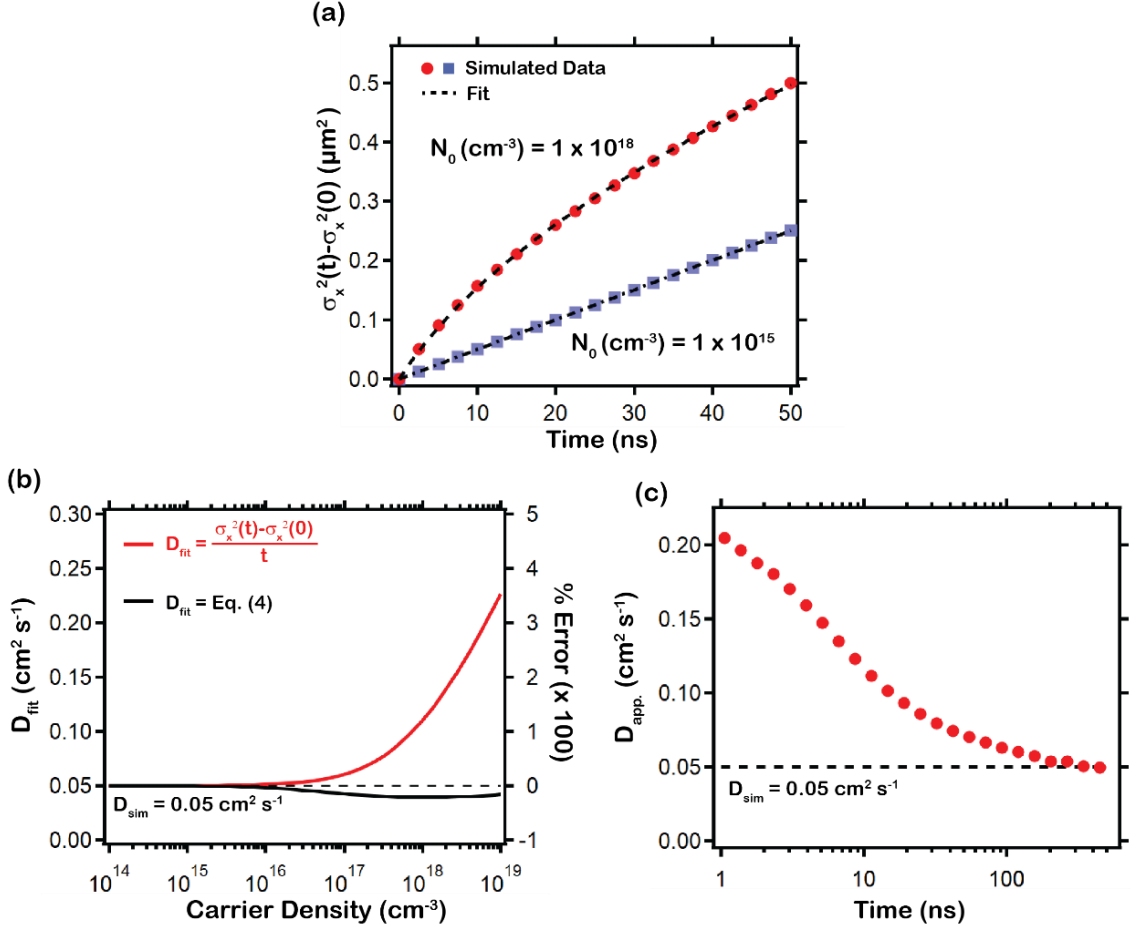

**Figure 3.** a) Fits to mean squared displacement curves at initial carrier densities ($N_0$) of $1\times10^{15}$ cm$^{-3}$ and $1\times10^{18}$ cm$^{-3}$ using the model, $D = [\sigma_x^2(t) - \sigma_x^2(0) - \int_0^t \frac{k_2 N(t')}{16\pi z} dt']/t$. b) $D_{fit}$ as a function of $N_0$ for a fixed $D_{sim}$ of 0.05 cm$^2$ s$^{-1}$ (dashed black line) using the linear equation, $D = [\sigma_x^2(t) - \sigma_x^2(0)]/t$, compared to the revised MSD model incorporating nonlinear recombination from a). c) Apparent diffusion coefficient ($D_{app}$) as a function of time calculated by taking the instantaneous slope, $D_{app.} = (MSD_{n+1} - MSD_n)/(t_{n+1} - t_n)$, of the curve for $1\times10^{18}$ cm$^{-3}$ in a). Domain size = 10 x 10 $\mu$m; $k_1$ = 5x10$^6$ s$^{-1}$; $k_2$ = 2x10$^{-10}$ cm$^3$ s$^{-1}$; and $D_{sim}$ = 0.05 cm$^2$ s$^{-1}$.

## IV. MSD MODEL FOR THIRD-ORDER RECOMBINATION KINETICS

From above, it appears that the MSD Model described in Eq. (2) serves as a valid approximation when the signal scales linearly with carrier density (cf. Figure 1) and the modified MSD Model in Eq. (4) can be utilized when the signal scales either linearly and/or quadratically with carrier density. Although our revised MSD model results in significantly less error for second-order recombination, most optical microscopy measurements are conducted in carrier density regimes where third-order (i.e. carrier-carrier Auger) processes are dominant (see Figure S3) and artificial broadening effects are expected to be even more significant. In order to further explore this regime, we simulate diffusion including third-order recombination as shown in the following Eq.

$$\frac{\partial N(\boldsymbol{u},t)}{\partial t} = \nabla \cdot \left(D_{ij}(\boldsymbol{u})\nabla N(\boldsymbol{u},t)\right) - k_1 N(\boldsymbol{u},\text{t}) - k_2 N^2(\boldsymbol{u},t) - k_3 N^3(\boldsymbol{u},t) \qquad (7),$$

where $k_3$ is the non-radiative, third-order recombination rate constant.

Figure 4a shows the simulated time-dependent MSD profiles with a fixed $D_{sim}$ = 0.05 cm$^2$ s$^{-1}$ over a 2 ns time-window for initial carrier densities of $1\times10^{18}$ cm$^{-3}$ and $1\times10^{19}$ cm$^{-3}$. The shorter time range is typical for microscopy experiments with an optical delay stage and was also chosen to emphasize the early time dynamics where third-order processes dominate at higher initial carrier densities (see Fig. S3). Similar to Figure 3, Figure 4 shows that third-order processes lead to a larger change in the MSD profile over a shorter time-span and must be considered in order to accurately extract $D$. Therefore, we expand upon our non-linear derivation presented in Eq. (4) and apply a similar methodology to develop a revised MSD Model which includes third-order processes.

$$D = \frac{\sigma_{x,2}^2(t) - \sigma_{x,2}^2(0) - \int_0^t \frac{tk_2 N(t')}{16\pi z} dt' - \int_0^t \frac{2k_3 N^2(t')}{36\pi^2 z^2 (\sigma_x^2(0) + 4Dt)} dt'}{t} \qquad (8)$$

Here, $N_{tot}(t)$ is the total time-dependent carrier population function

$$N_{tot}(t) = \frac{N_{tot}(0)\exp(-k_1 t)}{\sqrt{1 + \frac{2N_{tot}^2(0) t \exp(-2k_1 t)}{3\pi^2 \sigma_x^2(0)(\sigma_x^2(0) + 4Dt)}}} \qquad (9)$$

The last term in the numerator of Eq. (8) accounts for Auger recombination.

Figure 4a shows fits to the simulated data using Eqs. (8) and (9) and Figure 4b shows the range of initial carrier densities where Eq. (8) accurately (< 5% error) recovers $D_{sim}$ compared to the regions where significant error is introduced. These ranges are strongly dependent on the spot size, where the artificial broadening of the Gaussian profile becomes more pronounced for larger spots (see Figure S4c and Eqs. in SI). Although this data highlights the fitting error that arises when $D_{sim}$ = 0.05 cm$^2$ s$^{-1}$, it also suggests that the magnitude of error may be a function of the $D_{sim}$ values, as the diffusion coefficient controls the local carrier density and, hence, the dominant recombination pathways (see Figure S3). We find more significant error at lower $D_{sim}$ values, which could be expected as carriers are confined to smaller volumes leading to greater fractions of higher-order recombination. These results highlight the importance of conducting confocal measurements at low carrier densities (< 1 x 10$^{16}$ cm$^{-3}$), especially for perovskite samples where $D$ values typically range from 0.005 – 2 cm$^2$ s$^{-1}$ (see Figure S5).[33] Similar to Figure 3c, Figure 4c shows the apparent diffusion coefficient ($D_{app}$) spanning 10 ps to 500 ns for an initial carrier density of 1x10$^{19}$ cm$^{-3}$. $D_{app}$ varies over two orders of magnitude and can even reach values as high as ~100's cm$^2$ s$^{-1}$ (Fig. S6) for slighly larger spot sizes, which could appear as quasi-ballistic or ballistic transport. It is possible that the weak positive correlation (population size, N = 20) between carrier density and the reported diffusion coefficients (Figure S5), may be a result of systematic fitting error exacerbated at higher initial carrier densities and at early time scales. We note that this fitting error is still prevalent even when the non-radiative rate constant ($k_1$) is increased by three orders of magnitude to 1x10$^9$ s$^{-1}$ (see Figure S7),[33] indicating that the material quality (i.e. defect density) does not significantly impact these findings.

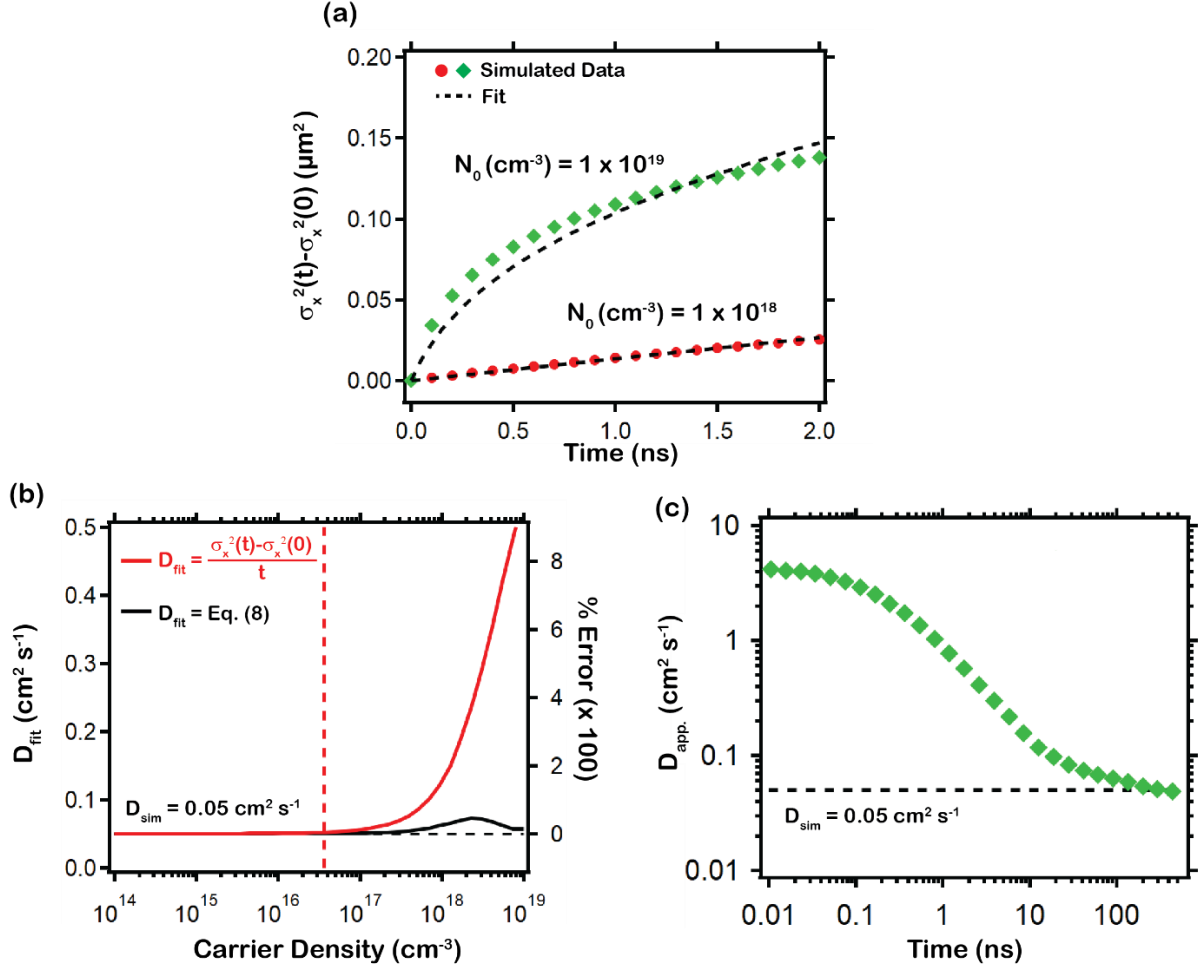

**Figure 4.** a) Fits to mean squared displacement curves at initial carrier densities ($N_0$) of $1\times10^{18}$ cm$^{-3}$ and $1\times10^{19}$ cm$^{-3}$ using the revised MSD model shown in Eq. (8). b) $D_{fit}$ as a function of $N_0$ for a fixed $D_{sim}$ of 0.05 cm$^2$ s$^{-1}$ (dashed black line) using the linear equation, $D = [\sigma_x^2(t) - \sigma_x^2(0)]/t$, compared to the nonlinear equation from a). c) Apparent diffusion coefficient ($D_{app.}$) as a function of time calculated by taking the instantaneous slope, $D_{app.} = (MSD_{n+1} - MSD_n)/(t_{n+1} - t_n)$, for $1\times10^{19}$ cm$^{-3}$ from a). Domain size = 5 x 5 $\mu$m; $k_1 = 5\times10^6$ s$^{-1}$; $k_2 = 2\times10^{-10}$ cm$^3$ s$^{-1}$; $k_3 = 1\times10^{-28}$ cm$^3$ s$^{-1}$; and $D_{sim} = 0.05$ cm$^2$ s$^{-1}$.

## V. IMPACT OF GRAIN SIZE AND GRAIN BOUNDARY BEHAVIOR ON DISTRIBUTION PROFILES

Thus far, we have only considered the evolution of the carrier distribution profile when the excitation spot size (with FWHM of 660 nm) is much smaller than the grain size (5 $\mu$m). Therefore, these simulations represent the carrier transport behavior before carriers reach the domain boundaries (i.e. grain boundaries). Given that grain size and grain boundary passivation are controllable material parameters which directly influence photovoltaic performance,[40-42] we next explore the role of domain size and boundary conditions (boundaries being reflective, transmissive, or serving as non-radiative recombination centers) on carrier transport properties. To better understand the impact of different domain sizes and domain boundary behavior on the apparent diffusion processes for a given excitation spot size, we perform simulations on 2, 3, and 5 $\mu$m sized square grains (~ 3x, 4.5x, and 7.5x the spot size, respectively) with reflective (i.e. no flux) grain boundaries. For this analysis, we specifically chose reflective boundary conditions as several recent studies have suggested that the grain boundaries in high-quality samples may act as impermeable barriers to carrier transport.[22,24,25,30,43,44] Figure 5a-b show the simulated PL maps at 50 ns for the different sized grains and the MSD plots, respectively. We observe a sublinear change in the MSD for the 2 and 3 $\mu$m grain sizes as a function of time, which is due to the grain boundaries acting as solid walls and the fixed carrier density profile decaying at a similar rate across the entire grain.[33] We highlight that this effect is not present in larger grains (i.e. > 5 $\mu$m) or single crystal samples and is especially prominent in smaller grains (i.e. < 3 $\mu$m), again for a $D_{sim}$ of 0.05 cm$^2$ s$^{-1}$. Importantly, this behavior appears similar to "sub-diffusive", nonequilibrium transport, or trap limited transport,[22,39] and might be enticing to fit with an $\alpha$ parameter. Such error in fitting can be avoided by studying the thermal activation of the diffusion process, the changes in raw signal intensity, as well as the spatiotemporal spectra, where a slow spectral redshift is often indicative of carriers undergoing multiple trapping or hopping mechanisms.[34,45] In addition to grain size, Figure 5c also shows the temporal evolution of the MSD for a 2 $\mu$m grain with quenching grain boundaries and a surface recombination velocity, $S$, of 600 cm s$^{-1}$ (see Figure S8 for PL profiles).[12,33] Here the change in the MSD stays nearly constant due to a non-zero carrier flux across the grain boundaries.

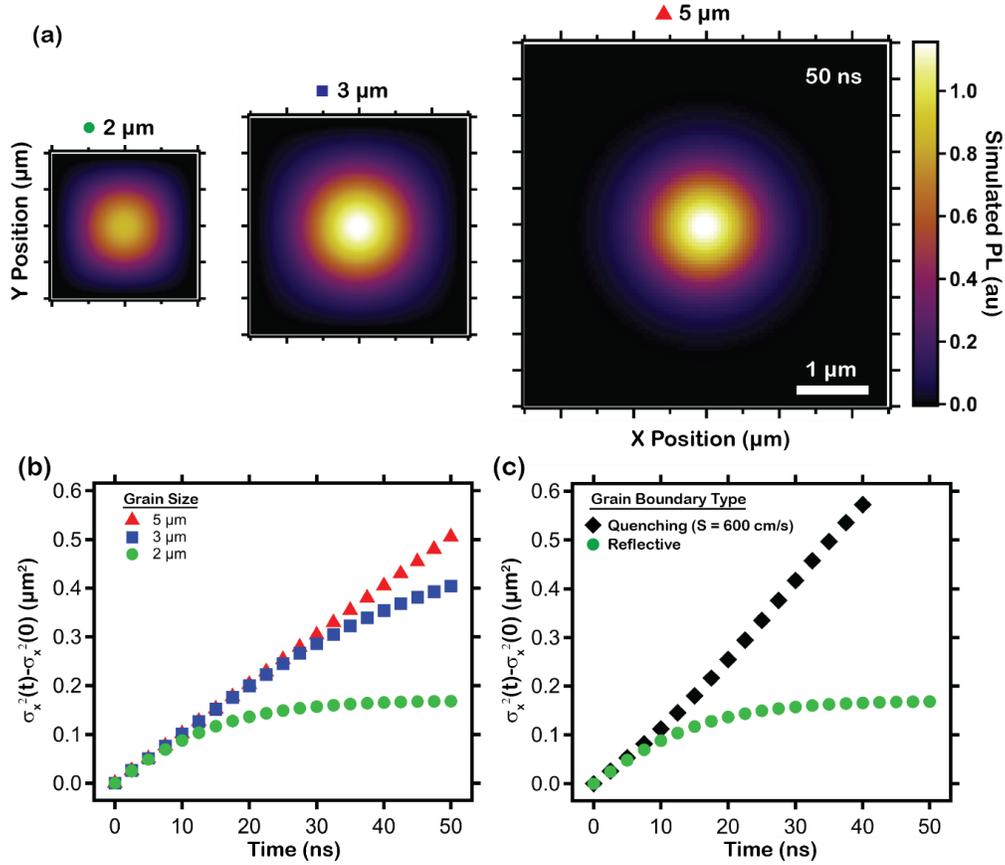

**Figure 5.** a) Simulated photoluminescence (PL) diffusion maps for grain sizes of 2, 3, and 5 $\mu$m at $t = 50$ ns with reflective grain boundary conditions. b) Mean squared displacement (MSD) plots of the simulations in (a) from 0 – 50 ns. c) MSD plots for 2 $\mu$m sized grains with quenching versus reflective boundary conditions. $N_0 = 1\times10^{16}$ cm$^{-3}$; $k_1 = 1\times10^6$ s$^{-1}$; $k_2 = 2\times10^{-10}$ cm$^3$s$^{-1}$; $k_3 = 1\times10^{-28}$ cm$^6$s$^{-1}$; $D_{sim} = 0.05$ cm$^2$ s$^{-1}$; $S$ = surface recombination velocity = 600 cm s$^{-1}$. Color and size scale bars are the same for all images.

## VI. IMPACT OF PHOTON RECYCLING ON DISTRIBUTION PROFILES

Finally, we consider how the process of photon emission followed by photon reabsorption (i.e. photon recycling) may impact the carrier distribution profile and whether this process could effectively compete with normal diffusive processes. In order to simulate diffusion processes along with photon recycling, we use a set of coupled partial differential equations as shown in Eqs. (10) and (11) to study carrier and photon dynamics, respectively:

$$\frac{\partial N(\boldsymbol{u},t)}{\partial t} = \nabla \cdot \left(D_{ij}(\boldsymbol{u})\nabla N(\boldsymbol{u},t)\right) + \sum \frac{c}{n(E)}\alpha(E)\gamma(E,\boldsymbol{u},t) - k_1 N(\boldsymbol{u},t) - k_2 N(\boldsymbol{u},t)^2 - k_3 N(\boldsymbol{u},t)^3$$

(10),

$$\frac{\partial \gamma(E,\boldsymbol{u},t)}{\partial t} = \nabla \cdot \left(D_{ij}^{\gamma}(E,\boldsymbol{u})\nabla \gamma(E,\boldsymbol{u},t)\right) + (1-P_{esc})k_2 N(\boldsymbol{u},t)^2 P(E) - \sum \frac{c}{n(E)}\alpha(E)\gamma(E,\boldsymbol{u},t)$$

(11),

where $c$ is the speed of light, $n(E)$ is the energy-dependent refractive index, $\alpha(E)$ is the absorption coefficient, $\gamma(E,\boldsymbol{u},t)$ is the spatiotemporal photon density, $D_{ij}^{\gamma}$ is the photon diffusion coefficient tensor which we approximate to be $8 \times 10^4$ cm$^2$ s$^{-1}$ (see SI for details), and $P(E)$ is the probability that a photon is emitted with a given energy. In this approach, we disregard polariton effects and consider the simplest model of the photon diffusion in the medium.[46,47]

Figure 6a shows the photoluminescence maps at 0 and 5 ns of a hypothetical perovskite thin film with and without photon recycling. We observe an increase in the the total PL for the case with photon recycling across the entire image, consistent with photon recycling leading to an increase in the average carrier density.[48] Figure 6b shows the normalized line profiles through the center of the maps at 0, 2.5 and 5 ns. Interestingly, although photon recycling leads to more carriers in each position, the spatial distribution of the carriers is not significantly effected by photon recycling (see Figure S9 for absolute values). This simulation is for a sample with an initial (t = 0 ns) internal photoluminescence quantum efficiency (PLQE) of 92%, where photon recycling is expected to have a large impact on the performance of photovoltaic devices.[48] We note that this model does predict enhanced transport with photon recycling along the depth of a thin (350 nm) film, which is consistent with simulated results from Ansari-Rad *et al.* (Fig. S10).[49,50] These results imply that, for the perovskite sample parameters used in this study, photon recycling does not significantly alter the line profiles and, therefore, does not need to be considered when extracting $D$. This is consistent with previous studies that have studied the uncoupled behavior of photons and photoexcited carriers.[36] We highlight that this observation only applies to materials with slow radiative decay rates and photon recycling leads to a marked change in the carrier profiles for materials that are continuously excited (see Fig. S11) or with fast radiative lifetimes, such as excitonic materials (see Fig. S12).

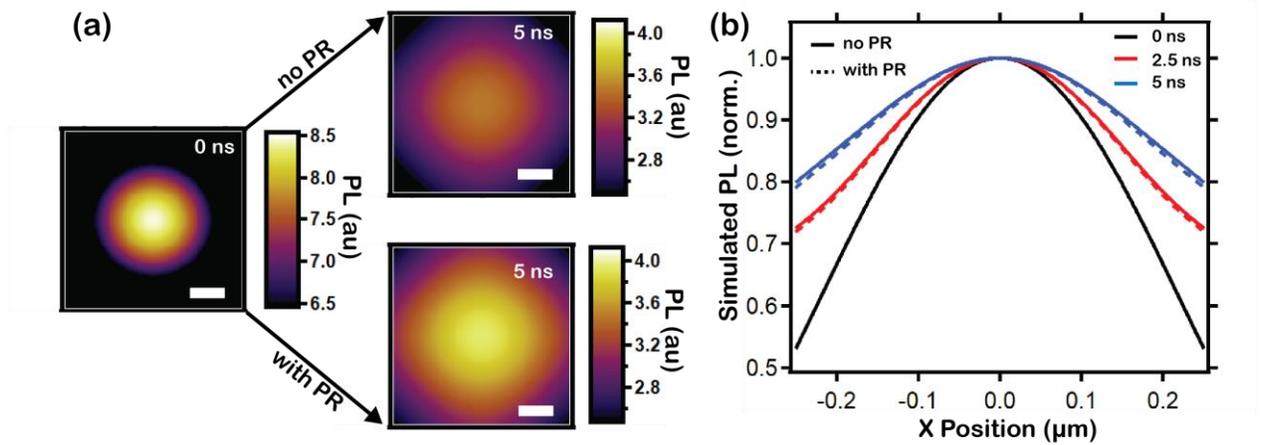

**Figure 6.** a) Simulated photoluminescence (PL) maps at temporal slices of t = 0 and 5 ns with and without photon recycling (PR). b) Normalized temporal PL line profiles along the *x*-direction at 0, 2.5, and 5 ns with (solid lines) and without PR (dashed lines). $N_0 = 1\times10^{17}$ cm$^{-3}$; Domain size = 500 x 500 nm; $k_1 = 5\times10^6$ s$^{-1}$; $k_2 = 2\times10^{-10}$ cm$^3$ s$^{-1}$; $k_3 = 1\times10^{-28}$ cm$^3$ s$^{-1}$; $D_{sim} = 0.05$ cm$^2$ s$^{-1}$; and $D^\gamma = 8\times10^4$ cm$^2$ s$^{-1}$. All images have the same color scale range and the scale bars are 1 µm.

## VII. DISCUSSION

The direct visualization of energy flow with optical microscopy offers the ability to study structure-function relationships in emerging semiconductors which possess disordered landscapes, inhomogeneities, and impurities. Accurate analysis of the diffusion coefficient in these materials requires an understanding of the dominant photophysical processes under various experimental conditions as well as the determination on how sample-specific morphology controls carrier transport. For example, we show several scenarios where the commonly deployed MSD model does not accurately recover the underlying transport properties. Specifically, Eq. (2) was intended for homogenous semiconductors where the measured signal scales linearly with the carrier density (i.e. excitonic molecular solids). For free carrier semiconductors that are homogeneous (i.e. single crystal or large-grain 3D perovskites) and where the measured signal non-linearly scales with the carrier density, we derived revised MSD Models captured by Eqs. (4) and (8), which introduce significantly less error over a wide range of carrier densities and material diffusion coefficients (cf. Figs. 3 and 4). Namely, we would expect the fitting of $D$ to be independent of carrier density up to the points where carrier-carrier scattering effects begin to dominate and/or band-edge degeneracy is broken.[51] We note that the general approaches described in this work not only apply to fluorescence imaging, but also to transient reflection, absorption, and scattering microscopies where the change in the measured signal is proportional to the dominant recombination mechanism.

Importantly, this study highlights that MSD models are not well suited for materials with microstructural features that are of similar size to the diffusion length, unless boundary behavior and its impact on carrier dynamics is well known.[24] This is especially important for samples that demonstrate grain boundary behavior similar to Fig. 5, where the slowing of the MSD expansion is due to carrier reflection off the grain boundaries. Here, a carrier may undergo multiple reflections before recombining, making the total distance it traveled large, despite its small displacement (i.e.

vector magnitude). Therefore, reported diffusion lengths should be denoted as vector or scalar quantities, depending on the definition used. The MSD Model, which monitors the normalized changes in the carrier distribution profiles, does not differentiate between these scenarios. Without careful attention to the impact of various structural features, carrier transport could be misinterpreted as sub-diffusive or trap-dominated. Therefore, in addition to analyzing the raw (un-normalized) data, additional complementary experiments such as hyperspectral imaging and temperature-dependent measurements should be performed to further refine kinetic models. In order to overcome some of these limitations, structural features such as grain boundaries can be directly measured with techniques such as scanning electron microscopy and be fed into numerical simulations.[24] In order to avoid significant fitting error, Table 1 summarizes the models discussed in this work which result in the lowest fitting error for specific recombination regimes and sample features.

| *Dominant Recombination Pathway* | *Model Resulting in Lowest Error* |
|---|---|
| $k_1N$ | $D = \dfrac{\sigma_x^2(t) - \sigma_x^2(0)}{(2)t}$ |
| $k_2N^2$ | $D = \dfrac{\sigma_x^2(t) - \sigma_x^2(0) - \int_0^t \dfrac{k_2 N(t')}{16\pi z} dt'}{t}$ |
| $k_3N^3$ | $D = \dfrac{\sigma_x^2(t) - \sigma_x^2(0) - \int_0^t \dfrac{k_2 N(t')}{16\pi z} dt' - \int_0^t \dfrac{2k_3 N^2(t')}{36\pi^2 z^2 (\sigma_x^2(0) + 4Dt)} dt'}{t}$ |
| *Structural Feature* | |
| **Grains and Grain Boundaries** | Partial Differential Equations with Sample-Specific Boundary Conditions |

Table 1. Mean-Squared-Displacement (MSD) Models that result in the lowest error for the fitted diffusion coefficient ($D$) when recombination is dominated by first-order (i.e. excitonic or trapping) $k_1N$, second-order (i.e. bimolecular) $k_2N^2$, and third-order (i.e. Auger) $k_3N^3$ processes. The "(2)" in the demoninator of the MSD Model for $k_1N$ should be removed if the signal intensity scales quadratically with the carrier density (i.e. PL in free carrier semiconductors). For materials with microstructural features on the scale of the diffusion length, partial differential equations with sample-specific boundary conditions should be used**.**

## VIII. CONCLUSION

We simulate the two-dimensional spatial and temporal dynamics of photoexcited energy carriers upon generation under a focused laser spot. We test the accuracy of the commonly deployed Mean Squared Displacement (MSD) Model, $D_{fit} = [\sigma_x^2(t) - \sigma_x^2(0)]/2t$ in extracting the diffusion coefficient, $D$, and find significant fitting error that increases at high initial carrier densities, larger spot sizes, low diffusion coefficients, and at early time scales (Figure S4). In order to reduce fitting error, we derive revised analytical MSD Models which can take into account both the second-order (i.e. bimolecular) and third-order (i.e. Auger) recombination processes that lead to artificial broadening of the carrier density profiles. When applied to a hypothetical perovskite sample with commonly reported physical parameters, we show a reduction in the fitting errors by up to two orders of magnitude. These approaches can be applied to nearly all excitonic and free carrier semiconductors and are accurate over a wide range of initial carrier densities, excitation spot sizes, intrinsic material diffusion coefficients, and collection time windows. We expect these models will find broad use in the accurate characterization of emerging semiconductors using optical microscopy and lead to better design and optimization of device architectures for electronic and optoelectronic applications. In addition, this work bolsters efforts in developing a fundamental understanding of other important transport properties such as carrier scattering times, velocities, mean free paths, and hopping rates.

**ACKNOWLEGEMENTS**

D.W.D., R.B., M.L., B.M., and V.B. acknowledge support for this project through the MIT-Tata GridEdge Solar Research Program, which is funded by the Tata Trusts. M.L. and R.B. acknowledge support from the National Science Foundation Graduate Research Fellowship under Grant No.

(1122374). D.W.D. thanks Stéphane Kéna-Cohen (École Polytechnique de Montréal) and Victor Burlakov (University of Oxford) for helpful discussions.

Supporting Information for

# Determination of Semiconductor Diffusion Coefficient by Optical Microscopy Measurements

*Dane W. deQuilettes,[1*] Roberto Brenes,[1,2] Madeleine Laitz,[1,2] Brandon T. Motes,[2] Mikhail M. Glazov,[3] Vladimir Bulović[1,2*]*

[1] Research Laboratory of Electronics, Massachusetts Institute of Technology, 77 Massachusetts Avenue, Cambridge, Massachusetts 02139, USA

[2] Department of Electrical Engineering and Computer Science, Massachusetts Institute of Technology, 77 Massachusetts Avenue, Cambridge, Massachusetts 02139, USA

[3] Ioffe Institute, 26 Politekhnicheskaya, 194021, Saint Petersburg, Russian Federation

*Corresponding Authors: danedeq@mit.edu; bulovic@mit.edu

## S1. Derivation of Mean Squared Displacement (MSD) Model

As outlined in the main article, this work is primarily concerned with the application of the standard MSD Model shown in Equation S1 to various material systems under different experimental conditions:

$$D = \frac{\sigma_x^2(t) - \sigma_x^2(0)}{2t} \tag{S1}$$

In order to specifically address the debated topics of when the model is appropriate to use and also which coefficients should be included in the denominator,[1] we outline the derivation below. We begin with the partial differential equation (PDE) which includes first-order recombination.

$$\frac{\partial N(\boldsymbol{u},t)}{\partial t} = \nabla \cdot (D_{ij}(N,\boldsymbol{u},t) \nabla N(\boldsymbol{u},t)) - k_{tot} N(\boldsymbol{u},t) \tag{S2}$$

We assume that the material is homogenous on the length scale of the diffusion length and therefore the diffusion coefficient is 1) isotropic (i.e. $D_{ij} = 0$ when $i \neq j$; $D_{ij} = D$ when $i = j$), 2) independent of time, 3) independent of carrier density —photoexcitation density is sufficiently low at $t = 0$, and 4) that the diffusion along different axes is uncorrelated (i.e. the particle has no memory of the direction of its last hop) and therefore we can only consider the diffusion along one dimension:

$$\frac{\partial N(x,t)}{\partial t} = D \frac{\partial^2}{\partial x^2} N(x,t) - k_{tot} N(x,t) \tag{S3}$$

We consider the initial condition ($t = 0$) where the initial carrier density profile is the Delta function at the origin, $\delta(x_0)$. Next, we know that the photoexcited carrier will spread out uniformly from the point of excitation becoming both wider and shallower and make an educated guess of a prototypical solution.[1]

$$N(x,t)|_{\delta(t=0)} = t^{-\beta} F(\eta) \tag{S4}$$

$$\eta = \frac{x^2}{4Dt} \tag{S5}$$

Here, $t^{-\beta}$ is the size-factor that decreases the carrier density amplitude as a function of time when $t^{-\beta} > F(\eta)$, and $F(\eta)$ is the shape-factor that dictates the curve's profile. The powers of both $x$ and $t$ are determined based on (S3), where $x$ is associated with a second-order derivative and $t$ as a first-order derivative. The introduction of $D$ into $\eta$ makes the ratio dimensionless. According to this definition, we will solve for the constant, $\beta$, and the function, $\eta$.

We calculate the derivatives of this prototypical solution using the product rule where appropriate:

$$\frac{\partial N}{\partial t} = -\beta t^{-\beta-1} F + t^{-\beta} \frac{dF}{d\eta} \frac{\partial \eta}{\partial t}$$

$$= -\beta t^{-\beta-1} F - \eta t^{-\beta-1} \frac{dF}{d\eta} \tag{S6}$$

$$\frac{\partial N}{\partial x} = t^{-\beta} \frac{dF}{d\eta} \frac{\partial \eta}{\partial x}$$

$$= \frac{x t^{-\beta-1}}{2D} \frac{dF}{d\eta} \tag{S7}$$

$$\frac{\partial^2 N}{\partial x^2} = \frac{t^{-\beta-1}}{2D} \frac{dF}{d\eta} + \frac{x t^{-\beta-1}}{2D} \frac{d^2 F}{d\eta^2} \frac{\partial \eta}{\partial x}$$

$$= \frac{\partial^2 N}{\partial x^2} = \frac{t^{-\beta-1}}{2D} \frac{dF}{d\eta} + \frac{x^2 t^{-\beta-1}}{4D^2 t} \frac{d^2 F}{d\eta^2}$$

$$= \frac{t^{-\beta-1}}{2D} \frac{dF}{d\eta} + \eta \frac{t^{-\beta-1}}{D} \frac{d^2 F}{d\eta^2} \tag{S8}$$

Substituting (S6) and (S8) into the PDE (S3) in the case where $k_{tot} = 0$:

$$-\beta t^{-\beta-1}F - \eta t^{-\beta-1}\frac{dF}{d\eta} = D\left(\frac{t^{-\beta-1}}{2D}\frac{dF}{d\eta} + \eta\frac{t^{-\beta-1}}{D}\frac{d^2F}{d\eta^2}\right) \tag{S9}$$

We factor and cancel terms to simplify Equation S9 as much as possible:

$$t^{-\beta-1}\left(-\beta F - \eta\frac{dF}{d\eta}\right) = t^{-\beta-1}\left(\frac{1}{2}\frac{dF}{d\eta} + \eta\frac{d^2F}{d\eta^2}\right) \tag{S10}$$

$$-\beta F - \eta\frac{dF}{d\eta} = \frac{1}{2}\frac{dF}{d\eta} + \eta\frac{d^2F}{d\eta^2} \tag{S11}$$

$$-\beta F - \frac{1}{2}\frac{dF}{d\eta} = \eta\left(\frac{dF}{d\eta} + \frac{d^2F}{d\eta^2}\right) \tag{S12}$$

$$-\frac{1}{2}\left(\frac{dF}{d\eta} + 2\beta F\right) = \eta\frac{d}{d\eta}\left(\frac{dF}{d\eta} + F\right) \tag{S13}$$

If we set $\beta = \frac{1}{2}$, then the PDE becomes a simple ODE

$$-\frac{1}{2}\left(\frac{dF}{d\eta} + F\right) = \eta\frac{d}{d\eta}\left(\frac{dF}{d\eta} + F\right) \tag{S14}$$

$$\frac{dF}{d\eta} + F = 0 \tag{S15},$$

which can be solved analytically:

$$F(\eta) = Ae^{-\eta} \tag{S16}$$

Substituting the solution into our prototypical solution from Equation S4.

$$N(x,t)|_{\delta(t=0)} = t^{-\beta}F(\eta) = At^{-\beta}e^{-x^2/4Dt} \tag{S17}$$

Where, again, we set $\beta = \frac{1}{2}$.

$$N(x,t)|_{\delta(t=0)} = At^{-1/2}e^{-x^2/4Dt} \tag{S18}$$

We determine the normalization factor, A, which is the integral of a Gaussian function with respect to x.

$$\int_{-\infty}^{\infty} A t^{-1/2} e^{-x^2/4Dt} \, dx = 1 \tag{S19}$$

$$A = \frac{1}{\sqrt{4\pi D}} \tag{S20}$$

Putting this all together, we arrive at the solution for the carrier density as a function of position and time. Importantly, the probability of finding an energy carrier at a position of x after a time of t is a Gaussian function of the form:

$$\boldsymbol{N(x,t)|_{\delta(t=0)} = G(x,t) = \frac{1}{\sqrt{4\pi Dt}} e^{-x^2/4Dt}} \tag{S21}$$

Including the linear recombination term in Equation S3 the equation becomes.

$$N(x,t)|_{\delta(t=0)} = G(x,t) = e^{-k_{tot}t} \frac{1}{\sqrt{4\pi Dt}} e^{-x^2/4Dt} \tag{S22}$$

The exponential prefactor with $k_{tot}$ strongly impacts the recombination rate, but is independent of x. Therefore, all positions across the carrier density profile decrease the same relative amount leading to no impact on the spatial broadening of the distribution. Taking this into account, the measured signal intensity is often normalized to isolate the expansion of the carrier density distribution.

**S1.1. Convolution with Gaussian Initial Condition**

Next, we consider the initial condition which, for a confocal microscopy experiment, is an Airy disk that can be approximated as a Gaussian function

$$N(x,0) = N_0 e^{-x^2/2\sigma_x^2} \tag{S23},$$

where $N_0$ is the initial carrier density [cm$^{-3}$] over the Gaussian area and film depth and therefore does not have a common Gaussian normalization prefactor of the form $(2\pi\sigma_x^2)^{-1}$. $\sigma_x^2$ is the variance and can formally be defined as

$$\sigma_x^2(t) = \frac{\int x^2 N(x,t)dx}{\int N(x,t)dx} \tag{S24}.$$

The time evolution of the carrier density is therefore the convolution of the initial carrier density with the evolving Gaussian profile (S22) is

$$N(x,t) = N(x,0) * G(x,t) = \frac{N_0 e^{-k_{tot}t}}{\sqrt{4\pi Dt}} \int_{-\infty}^{\infty} e^{-x^2/2\sigma_x^2} e^{-x^2/4Dt} dx \tag{S25}.$$

Conveniently, the convolution of two Gaussians functions with variances of $\sigma_A^2$ and $\sigma_B^2$ is $\sigma_A^2 + \sigma_B^2$.

$$N(x,t) = \frac{N_0 e^{-k_{tot}t}}{\sqrt{4\pi Dt}} e^{-\left(x^2/(2\sigma_x^2+4Dt)\right)} \tag{S26}$$

The variance, which is the excited carrier distribution ($\sigma_{N,x}^2$), is

$$2\sigma_{N,x}^2(t) = 2\sigma_x^2 + 4Dt \tag{S27}$$

and is often written as the mean squared displacement (MSD)

$$MSD = \langle x(t)^2 \rangle - \langle x(0)^2 \rangle = \sigma_{N,x}^2(t) - \sigma_x^2 = 2Dt \tag{S28}$$

Rearranging for $D$

$$\boldsymbol{D = \frac{\sigma_{N,x}^2(t) - \sigma_x^2}{2t}} \tag{S29}$$

Oftentimes, we do not have direct access to the local carrier density or the initial condition and measure a spectroscopic signal (i.e. such as photoluminescence or transient absorption) to gain insight into these distributions. As pointed out by others,[2] the variance of the measured signal intensity ($\sigma_{int}^2$) is the variance of the carrier distribution profile convolved with the point spread function ($\sigma_{PSF}^2$) of the measurement setup.

$$\sigma_{int}^2(t) = \sigma_{N,x}^2(t) + \sigma_{PSF}^2 = \sigma_x^2 + 2Dt + \sigma_{PSF}^2 \tag{S30}$$

The MSD is

$$\sigma_{int}^2(t) - \sigma_{int}^2(0) = \sigma_x^2 + 2Dt + \sigma_{PSF}^2 - (\sigma_x^2 + 2D(0) + \sigma_{PSF}^2) \tag{S31}$$

$$\sigma_{int}^2(t) - \sigma_{int}^2(0) = 2Dt \tag{S32}$$

$$\boldsymbol{D} = \frac{\sigma_{int}^2(t) - \sigma_{int}^2(0)}{2t} \tag{S33}$$

Throughout the main article and supporting information, we use $\sigma_x^2$ for simplicity, which often refers to the measured or simulated signal variance along the *x*-direction.

**S2. MSD Model for I(*t*) ~ $N^2$ when $\frac{dN}{dt} \sim -kN$**

Following the general approach described above, if the measured signal intensity (i.e. I(*t*) ~ PL) scales as $N^2$, then Equation S22 becomes

$$I(t) \sim N^2(x,t) = e^{-2k_{tot}t} \frac{1}{4\pi Dt} e^{-x^2/2Dt} \tag{S34}$$

Convolving with the initial condition defined in Equation S23:

$$= e^{-2k_{tot}t} \frac{N_0}{4\pi Dt} e^{-\left(x^2/(2\sigma_{x,2}^2 + 2Dt)\right)} \tag{S35}$$

Here we use $\sigma_{x,2}^2$ to denote the variance of the measured being quadratically proportional to the carrier density, which can be defined as

$$\sigma_{x,2}^2(t) = \frac{\int x^2 N^2(x,t) d^2x}{\int N^2(x,t) d^2x} \tag{S36}.$$

The time-dependent variance is

$$2\sigma_{x,2}^2(t) = 2\sigma_{x,2}^2(0) + 2Dt \tag{S37},$$

which, following Equations S30-S33, can be simplified to

$$\boldsymbol{D} = \frac{\sigma_{x,2}^2(t) - \sigma_{x,2}^2(0)}{t} \tag{S38}.$$

Figure S2 shows the mean squared displacement over a time-span of 50 ns for the scenario where PL ~ $k_2 N^2$ at a low simulated initial carrier density of $N_0$ of 1 x $10^{15}$ cm$^{-3}$. Here, first-order recombination is the dominant decay mechanism. We apply Equation S29 to fit the simulated data and obtain a $D_{fit}$ of 0.025 cm$^2$ s$^{-1}$, which underestimates $D_{sim}$ (0.05 cm$^2$ s$^{-1}$) by a factor of 2. This 50% error primarily arises from the fact that Equation 29 was derived assuming the signal scales as $k_R N$. For many intrinsic, free carrier semiconductors, the PL is a second-order (bimolecular) process due to the recombination of free electrons with free holes (i.e PL ~ $k_2 N^2$).[2]

At low carrier densities, this error can be somewhat avoided using the slightly modified version as shown in Equation S38. Figure S2 shows the correct application of Equation S38 for the cases where PL ~ $kN^2$ and the kinetics are dominated by first-order (i.e. trapping) processes, which results in the correct extraction of a $D_{fit}$ value of 0.050 cm$^2$ s$^{-1}$. At higher carrier densities (> 1x$10^{17}$ cm$^{-3}$), Eq. S38 no longer captures the nonlinear broadening and the revised MSD Models must be used.

**S3. Analytical Solutions with Non-Linear Recombination and Diffusion**

Let us consider the diffusion of particles in two-dimensions in the presence of higher-order recombination processes, which can be described by the diffusion equation

$$\frac{\partial N(\boldsymbol{u},t)}{\partial t} = \nabla \cdot \left(D_{ij}(\boldsymbol{u}) \nabla N(\boldsymbol{u},t)\right) - k_1 N(\boldsymbol{u},t) - k_2 N^2(\boldsymbol{u},t) - k_3 N^3(\boldsymbol{u},t) \tag{S39},$$

where the spatial coordinates are represented by the vector $\boldsymbol{u} = (x, y)$, $N(\boldsymbol{u}, t)$ is the spatially and temporally-dependent carrier density, $D_{ij}(\boldsymbol{u})$ is the diffusion coefficient tensor, $\nabla$ is the gradient operator, $k_1$ is the non-radiative, first-order (monomolecular) recombination constant; $k_2$ is the second-order (bimolecular) recombination rate constant; and $k_3$ is the non-radiative, third-order (Auger) recombination rate constant.

*Self-Consistent Ansatz*

As a zero-order approximation we use the Gaussian ansatz for $N(\boldsymbol{x}, t)$ in the form

$$N(\boldsymbol{x}, t) = \frac{N_{tot}(0) e^{-k_1 t}}{\pi \sigma_x^2 (1 + \frac{4 D_{app} t^2}{\sigma_x^2}) \exp(-x^2/(\sigma_x^2 + 4Dt))} \tag{S40}$$

where $D_{app}$ is the apparent diffusion coefficient (generally, $D_{app}$ is a function of time, see below), the total number of particles is defined below.

$$N_{tot}(0) = N_0 \pi \sigma_x^2(0) z \tag{S41},$$

$N_0$ is the carrier density and $z$ is the film thickness.

We also introduce the averages of $N^2(\boldsymbol{u}, t)$ and $N^3(\boldsymbol{u}, t)$, respectively, as

$$N_{2,0} = 2\pi \int_0^\infty N^2(\boldsymbol{x}, t) x \, dx = \frac{N_{tot}^2(0)}{2\pi (\sigma_x^2 + 4Dt)} \tag{S42}$$

$$N_{3,0} = 2\pi \int_0^\infty N^3(\boldsymbol{x}, t) x \, dx = \frac{N_{tot}^3(0)}{3\pi^2 (\sigma_x^2 + 4Dt)^2} \tag{S43}$$

Integrating Eq. (S39) over the area and substituting Eqs. (S40-43) we arrive at the following ordinary differential equation (ODE) for the total density of the particles

$$\frac{dN_{tot}(t)}{dt} = -k_1 N_{tot}(t) - \frac{k_2 N_{tot}^2(0)}{2\pi (\sigma_x^2 + 4Dt)} - \frac{k_3 N_{tot}^3(0)}{3\pi^2 (\sigma_x^2 + 4Dt)^2} \tag{S44}.$$

**S3.1 Analytical Solution with Second-Order (Bimolecular) Recombination and Diffusion**

In the specific case when $k_3 = 0$, Kulig *et al.* derived the change in the apparent, or effective, diffusion coefficient ($D_{app}$) as a function of time for 2D materials where Auger processes (i.e. $R_A$

[cm² s⁻¹]) begin to dominate at higher excitation densities.[3] Here, we extend this derivation for the case of a material with appreciable film thickness ($z$) and with a bimolecular recombination rate constant that is volumetric (*i.e.* $k_2$ [cm³ s⁻¹]). Following steps of the derivation by Kulig *et al.*, we integrate the second-order, partial differential equation over the volume of the thin film to solve for the time-dependent, and spatially *independent*, carrier *population, N*(t),

$$N_{tot}(t) = \frac{N_{tot}(0)\exp(-k_1 t)}{1 + \frac{k_2 N_{tot}(0)}{8\pi D z}\exp\left(\frac{C\sigma_{x,2}^2(0)k_1}{4D}\right)\gamma(t)} \tag{S45}.$$

where $\gamma(t)$ is a summation of exponential integrals,

$$\gamma(t) = \text{Ei}\left(-\frac{C\sigma_{x,2}^2(0)k_1}{4D} - k_1 t\right) - \text{Ei}\left(-\frac{C\sigma_{x,2}^2(0)k_1}{4D}\right) \tag{S46}.$$

Here, *C* is a correction coefficient modifying the variance and can also be used as a free variable in the fitting algorithm. In our simulations, it consistently converged to 5.4 over the entire simulated carrier density range. This method of approximation can also be used to describe the apparent diffusion coefficient ($D_{app}$).[3]

$$\frac{dD_{app}(t)t}{dt} = D + \frac{k_2 N_{tot}(t)}{16\pi z} \tag{S47}$$

Here, *D* is the true diffusion coefficient. We note that for a finite time step, $dD_{app}(t)t = \sigma_x^2(t) - \sigma_x^2(0)$ from Equation S38. Making this substitution and solving for the mean squared displacement, we integrate and arrive at

$$\boldsymbol{\sigma_x^2(t) - \sigma_x^2(0) = Dt + \int_0^t \frac{k_2 N_{tot}(t')}{16\pi z} dt'} \tag{S48}$$

Using these equations, one can account for common errors introduced using the standard MSD Model, namely second-order, non-linear effects are taken into account. In addition, artificial broadening effects become more pronounced for larger spots sizes (cf. Figure 4) and, importantly, these non-linear equations also take into account the dependence on the initial excitation spot.

**S3.2 Analytical Solution with Third-Order (Auger) Recombination and Diffusion**

Next, we consider the case where the $k_3$-processes are the dominant. In this situation we neglect the terms with $k_1$ and $k_2$ to arrive at

$$\frac{dN(t)}{N^3(t)} = -\frac{k_3 dt}{3\pi^2(\sigma_x^2+4Dt)^2} \tag{S49},$$

Trivial integration yields

$$N_{tot}(t) = \frac{N_{tot}(0)\exp(-k_1 t)}{\sqrt{1+\frac{2N_{tot}^2(0)t\exp(-2k_1 t)}{3\pi^2 \sigma_x^2(0)(\sigma_x^2(0)+4Dt)}}} \tag{S50}$$

In the general case where all three types of recombination processes are present, Eq. (S39) can be readily solved numerically. Next, we determine the functional form of the effective diffusion coefficient. First, we calculate the second moment of the particles distribution function

$$\rho_2(t) = 2\pi \int_0^\infty N^3(x,t)x^3 dx \equiv (\sigma_x^2 + 4Dt)N_0 \tag{S51}$$

We integrate with the weight $x^2$ over the area and obtain the equation for $\rho_2$:

$$\frac{d\rho_2(t)}{dt} = -4DN(t) - k_1\rho_2(t) - \frac{k_2 N^2(t)}{4\pi} - \frac{k_3 N^3(t)}{9\pi^2(\sigma_x^2+4Dt)} \tag{S52},$$

The apparent diffusion coefficient $D_{app}$ depends on time provided that non-linearities are sufficiently strong. Making use of Eq. (S44) we obtain

$$\frac{dD_{app}(t)t}{dt} = D + \frac{k_2 N(t)}{16\pi z} + \frac{2k_3 N^2(t)}{36\pi^2 z^2(\sigma_x^2(0)+4Dt)} \tag{S53}$$

We note that for a finite time step, $dD_{app}(t)t = \sigma_x^2(t) - \sigma_x^2(0)$. Making this substitution and solving for the mean squared displacement, we integrate Equation S53 to arrive at

$$\boldsymbol{\sigma_x^2(t) - \sigma_x^2(0) = Dt + \int_0^t \frac{k_2 N(t')}{16\pi z}dt' + \int_0^t \frac{2k_3 N^2(t')}{36\pi^2 z^2(\sigma_x^2(0)+4Dt)}dt'} \tag{S54}$$

A correction factor, $C$, can be introduced as a prefactor in front of the variance in Eqs. S50 and S54 to obtain fits with lower error. For our simulations with third-order recombination and those reported in the main article, we find that a $C$ value fixed at 4.6 led to the lowest error over the entire simulated carrier density range.

## S4. Recombination/Diffusion Model

As described in the main article, Equation S39 describes the spatiotemporal carrier density in two dimensions with the inclusion of higher-order recombination terms, where the carrier diffusivity is assumed to be the same for all charge carrier species. For our simulations, $D = 0.05$ cm$^2$s$^{-1}$, $k_1 = 1\times10^6$ s$^{-1}$, and $k_2^{int} = 2\times10^{-10}$ cm$^3$ s$^{-1}$, and $k_3 = 1\times10^{-28}$ cm$^6$ s$^{-1}$ unless otherwise stated. We also assume that diffusion along the $z$-dimension can be ignored for sufficiently thin, homogeneous samples. We note that many polycrystalline films do possess heterogeneity over the length scale of the film thickness, which can more accurately be described by a 3-dimensional partial differential equation.[3]

### S4.1. Gaussian Initial Condition

For the majority of the simulations in this work, we approximate the excitation spot with a Gaussian function defined in Equation S23, but in two-dimensions. For the simulation, we input the experimentally measured laser excitation profile parameters ($\sigma_x = \sigma_y = 314$ nm) and center the excitation source at the origin ($x = y = 0$).

### S4.2. Airy Disk Initial Condition

We also consider the excitation profile for a diffraction-limit spot in confocal microscopy, which is often described by an Airy disk. The functional form of an elliptical Airy disk excitation profile is given by:

$$G(\mathbf{r},0) = \frac{N_0}{I_0}\left(\frac{2J_1(1.22\pi r)}{1.22\pi r}\right)^2 \tag{S55}$$

Where $J_1$ is the Bessel function of the first kind, $\mathbf{r} = \sqrt{\left(\frac{x}{x_0}\right)^2 + \left(\frac{y}{y_0}\right)^2}$, $x_0$ is the x-coordinate along the x-axis where the first zero occurs, and $y_0$ is the y-coordinate along the y-axis where the first zero occurs, $I_0 = \iint_A B(\mathbf{r})dxdy$ where $B(\mathbf{r}) = \left(\frac{2J_1(1.22\pi r)}{1.22\pi r}\right)^2$ and $A$ is the domain area of the simulation.

### S4.3. Boundary Conditions and Photoluminescence or Transient Absorption Signal

For simulating the time-resolved PL maps and profiles we consider both no flux (i.e. reflective) boundary conditions in Equation S56 as well as quenching boundary conditions in Equation S57 with a surface recombination velocity (S) of 600 cm s$^{-1}$.[4]

$$\left.\frac{\partial N(\mathbf{u},t)}{\partial x}\right|_\gamma = \left.\frac{\partial N(\mathbf{u},t)}{\partial y}\right|_\gamma = 0 \tag{S56}$$

$$\left.\frac{\partial N(\mathbf{u},t)}{\partial x}\right|_\gamma = \left.\frac{\partial N(\mathbf{u},t)}{\partial y}\right|_\gamma = -\frac{S}{D}N(\boldsymbol{\gamma},t) \tag{S57}$$

where $\boldsymbol{\gamma}$ is the vector that denotes the spatial coordinates of the grain boundaries. The solutions to these equations (i.e. $N(\mathbf{u},t)$) are evaluated using a time-domain, finite-element analysis in the MATLAB PDE toolbox. For the radiative, first-order recombination processes described in Figure 1 and S1 the PL is calculated using

$$PL(t) = k_{rad}N(\mathbf{u},t) \tag{S58}$$

For the rest of the simulations throughout the work, the PL is the external radiative second-order recombination rate constant multiplied by the square of the time-dependent carrier population

$$PL(t) = k_2^{ext} N(\boldsymbol{u}, t)^2 \tag{S59}$$

These simulations have also been performed to monitor the transient absorption signal (TA ~ N(**u**,t)), which probes both radiative and non-radiative decay pathways,

$$\frac{dTA(t)}{dt} = -k_1 N(\boldsymbol{u}, t) - k_2^{int} N(\boldsymbol{u}, t)^2 - k_3 N(\boldsymbol{u}, t)^3 \tag{S60}$$

### S5. Photon Recycling Effects on Carrier Profiles

As discussed in the main article, we can explicitly include the effects of photon recycling in our simulations by using a set of coupled partial differential equations shown in Equations S61 and S62:

$$\frac{\partial N(\boldsymbol{u},t)}{\partial t} = \nabla \cdot \left(D_{ij}(\boldsymbol{u}) \nabla N(\boldsymbol{u},t)\right) + \sum \frac{c}{n(E)} \alpha(E) \gamma(E, \boldsymbol{u}, t) - k_1 N(\boldsymbol{u}, t) - k_2 N(\boldsymbol{u}, t)^2 - k_3 N(\boldsymbol{u}, t)^3$$

$$\tag{S61}$$

$$\frac{\partial \gamma(E,\boldsymbol{u},t)}{\partial t} = \nabla \cdot \left(D_{ij}(E, \boldsymbol{u}) \nabla \gamma(E, \boldsymbol{u}, t)\right) + (1 - P_{esc}) k_2 N(\boldsymbol{u}, t)^2 P(E) - \sum \frac{c}{n(E)} \alpha(E) \gamma(E, \boldsymbol{u}, t)$$

$$\tag{S62},$$

where $c$ is the speed of light, $n(E)$ is the energy-dependent refractive index, $\alpha(E)$ is the absorption coefficient, $\gamma(E, \boldsymbol{u}, t)$ is the spatiotemporal photon density, $P_{esc}$ is the average photon probability of escape from the film, and $P(E)$ is the probability that a photon is emitted with a given energy. Partially reflecting boundary conditions were used at the grain boundaries for $\gamma(E, \boldsymbol{u}, t)$ as previously described by Ansari-Rad *et al.*.[4]

We simplify our model by summing over all energies for the photon distribution and use an absorption coefficient $\alpha_e$ that is determined by spectrally weighting $\alpha(E)$ with $P(E)$ determined from experimental photoluminescence spectra. For a $CH_3NH_3PbI_3$ film, we obtained an $\alpha_e$ of $1.4 \times 10^4$ cm$^{-1}$ which was used for the 2D simulations. For the 1D simulations in Figure S10, we used $\alpha_e = 5 \times 10^4$ cm$^{-1}$ for consistency with the simulations done by Ansari-Rad et al.,[4] though we note that this value has no experimental basis. The average photon probability of escape $P_{esc}$ used for all the simulations was 7%, as calculated by Pazos-Outón et al.[5]

As described in the main article, the normalized photon profiles in Figure 6b show only small effects when photon recycling is considered. The line profiles actually show a small amount of narrowing when photon recycling is included, which can be explained by local differences in carrier density. Specifically, although the carrier density is enhanced throughout the whole profile, it is enhanced to a greater extent in areas where the carrier density is already high (more photon recycling at higher carrier densities), increasing the peak of the profile. When normalized, since the peak is higher with photon recycling than without photon recycling, this makes the profile appear slightly more narrow.

To determine if photon recycling effects can be observed under more extreme simulation parameters, we perform the same simulation, this time without carrier diffusion. Figure S11a shows small differences in the PL profiles with and without photon recycling especially at the tails of the PL profile. We note that the tails are several orders of magnitude below the peak of the profile, which suggests that these differences would be difficult to measure experimentally. Furthermore, at later decay times, the signal has decreased by several orders of magnitude relative to the initial signal, at which point the differences with and without photon recycling would be even more difficult to measure. Therefore, the effects of photon recycling in modifying the time-dependent carrier density profile are negligible for typical 3D perovskite films.

These results do not contradict recent measurements performed by Pazos-Outón *et al.*,[6] where they observed clear spectral effects of photon recycling far away from the excitation spot and which were conducted under a continuous generation source (i.e. CW laser source). For example, Figure S11b-c shows a nearly identical simulation as Fig. S11a, but using a constant generation source, and either excluding or including carrier diffusion. Here, we observe that photon recycling increases the tails of the PL profile, but this time the signal is only a few orders of magnitude below the signal peak. This suggests that the differences with and without photon recycling can be measured experimentally with a suitable setup such as lock-in detection.

Finally, we consider if there are any time-dependent scenarios where photon recycling leads to appreciable changes in the carrier density profiles. Figure S12 shows the impact of photon recycling in an excitonic material with a short radiative lifetime of 5 ns and with no carrier diffusion. In this time-dependent simulation, we observe clear tails in the PL signal that could affect the fitted Gaussian variance when using a MSD model. For example, the tails of the distribution are about an order of magnitude below the peak of the signal after 1 ns into the simulation. We highlight that at this point in the decay, there is still appreciable signal that many microscopy techniques could still measure. In such cases, where the radiative lifetime is comparable to the measurement window, the MSD Models discussed in this work could overestimate the diffusion coefficient. Therefore, more rigorous models that incorporate photon recycling, such as the PDE's used in this work, could be utilized to accurately recover the material diffusion coefficient.

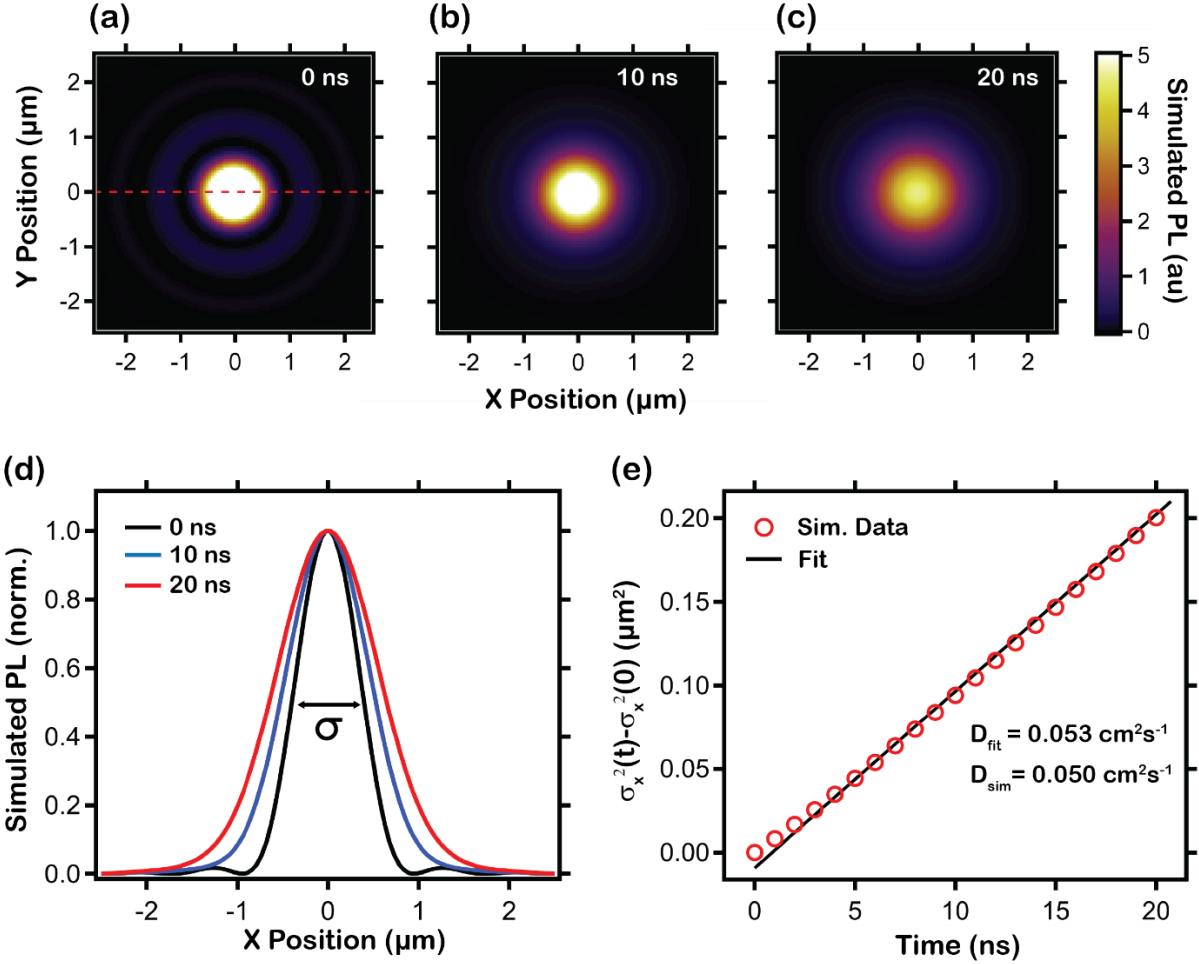

**Figure S1.** a-c) Simulated photoluminescence (PL) diffusion maps at time ($t$) intervals of $t = 0$, 10, and 20 ns. The initial excitation condition is an Airy disk and the dynamics follow first-order ($k_{tot}$) recombination kinetics (Domain size = 5 x 5 $\mu$m; $N_0$ = 1x10$^{16}$ cm$^{-3}$; $k_{tot}$ = 1x10$^6$ s$^{-1}$; $D_{sim}$ = 0.05 cm$^2$ s$^{-1}$; PL ~ $k_R N$, where $k_{tot}$ ~ $k_R$). d) Temporal line profiles along the $x$-direction, indicated by the red dashed line shown in (a). e) Gaussian variance as function of time along with a fit using the Mean Squared Displacement Model, where $D_{fit} = \frac{\sigma_x^2(t) - \sigma_x^2(0)}{2t}$. The change in the initial condition introduces a 6% error in the diffusion coefficient fit ($D_{fit}$ = 0.053 cm$^2$ s$^{-1}$) compared to the pre-defined $D_{sim}$ value input into the partial differential equation.

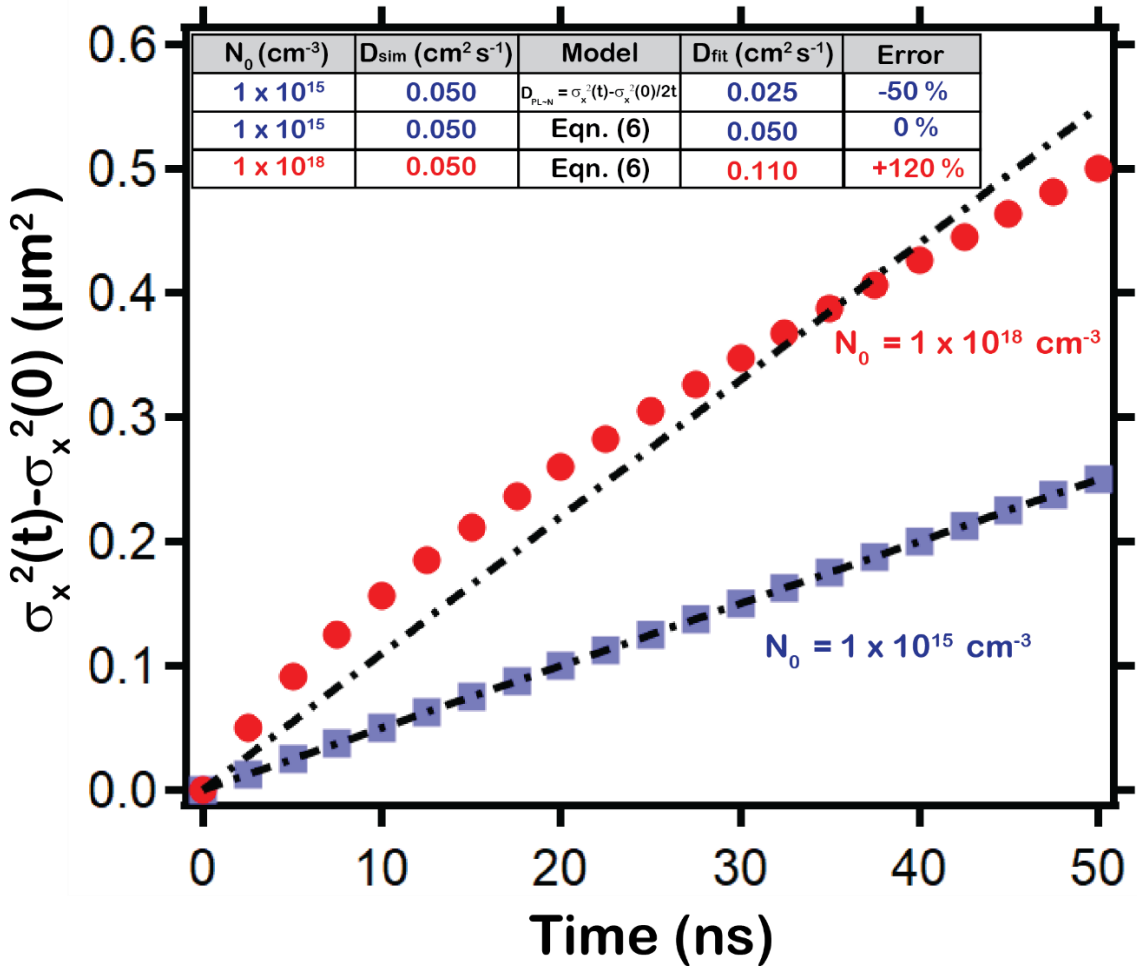

**Figure S2.** Fits to simulated photoluminescence (PL) data using Mean Square Displacement (MSD) Models ($i.e. D_{PL \sim N} = [\sigma^2(t) - \sigma^2(0)]/(2t)$) and ($D_{PL \sim N^2} = [\sigma^2(t) - \sigma^2(0)]/(t),$). Domain size = 2.5 x 2.5 $\mu$m, $k_1$ = 5x10$^6$ s$^{-1}$, and $k_2$ = 2x10$^{-10}$ cm$^3$ s$^{-1}$.

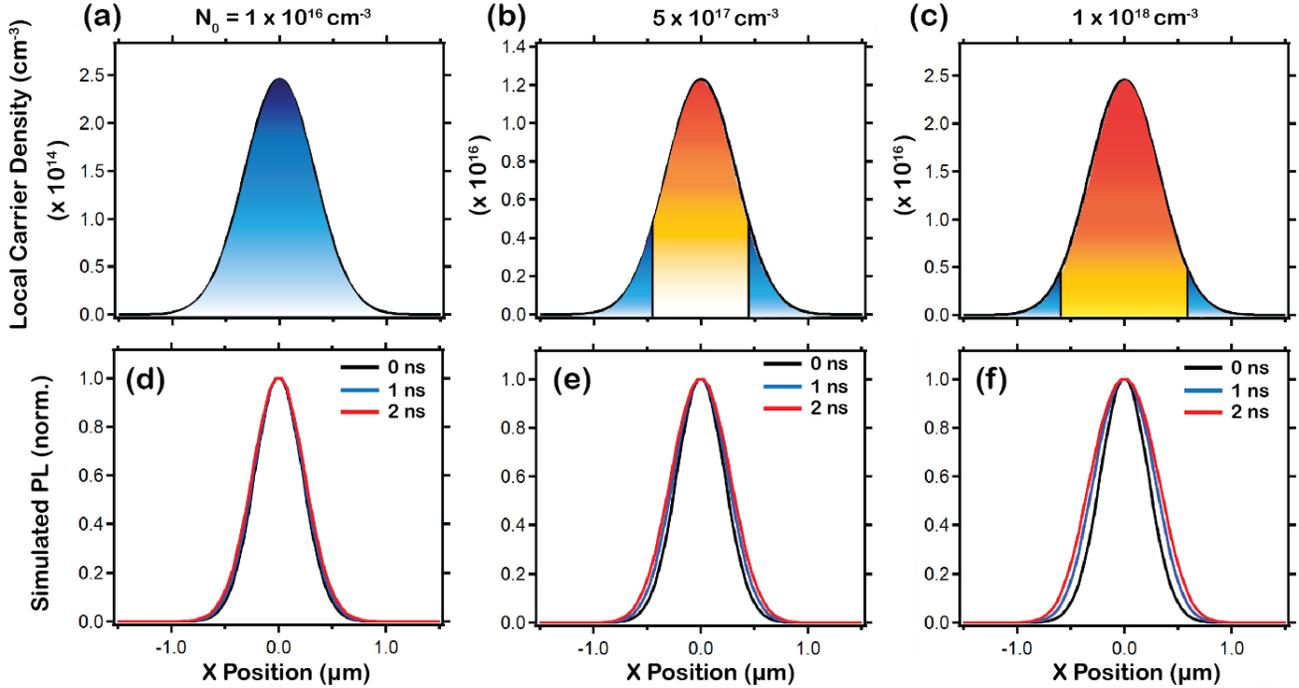

**Figure S3.** a-c) Increasing initial carrier density at $t = 0$ ($N_0$ is the integrated carrier density), showing the regions under the Gaussian excitation spot that are dominated (> 50% of the total recombination rate) by first-order ($k_1$, blue) versus higher-order ($k_2$ and $k_3$, red/orange) recombination for $N_0 = 1 \times 10^{16}$ cm$^{-3}$, $5 \times 10^{17}$ cm$^{-3}$, and $1 \times 10^{18}$ cm$^{-3}$, respectively. d-f) Time-dependent photoluminescence (PL) line profiles along the $x$-direction at $t = 0$, 1, and 2 ns for the three different $N_0$'s. As the initial carrier density increases, the area of the Gaussian profile dominated by the $k_2$ and $k_3$ terms extends further out to the tails of the Gaussian distribution.

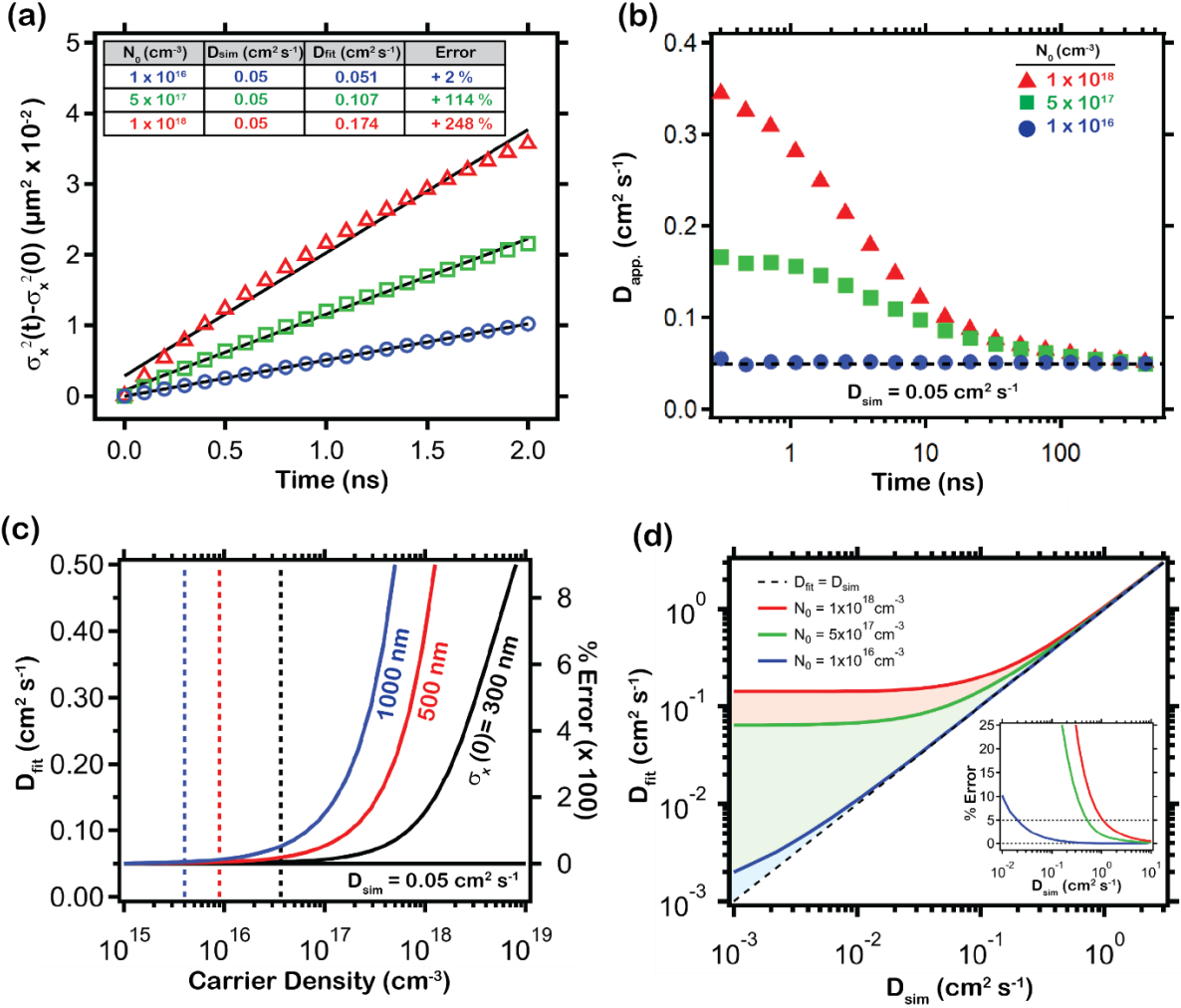

**Figure S4.** a) Mean squared displacement (MSD) plots and fits to the profiles in Figure 3d-f in the main article using the MSD Model, $D_{fit} = [\sigma_x^2(t) - \sigma_x^2(0)]/t$, for $N_{0,total} = 1 \times 10^{16}$ cm$^{-3}$ (blue circles), $5 \times 10^{17}$ cm$^{-3}$ (green squares), $1 \times 10^{18}$ cm$^{-3}$ (red triangles) over a 2 ns time window. b) Identical plot from (a) over a 500 ns time-window. c) Apparent diffusion coefficient ($D_{app}$) as a function of time calculated by taking the instantaneous slope $\left(i.e. \frac{(MSD_{n+1} - MSD_n)}{(t_{n+1} - t_n)}\right)$ for the three carrier densities. The slopes do not converge until ~ 500 ns. The differences in slopes for a given time interval are different due to the differences in the instantaneous carrier densities and, hence, dominant kinetic pathways. c) $D_{fit}$ as a function of the initial carrier density ($N_0$) for a fixed $D_{sim}$ of 0.05 cm$^2$ s$^{-1}$ for spot sizes with $\sigma_x = 300$ (solid black line), 500 (solid red line), and 1000 nm (solid

blue line). The colored dashed lines delineate the transition to greater than 5% fitting error for the corresponding spot size. d) $D_{fit}$ as a function of $D_{sim}$ for the three $N_0$ values, shaded regions represent the error from the true line $D_{fit} = D_{sim}$. (inset) Error percentages over a subset of the full $D_{sim}$ range, where $\% \, Error = 100\, \% \cdot |D_{fit} - D_{sim}|/D_{sim}$.

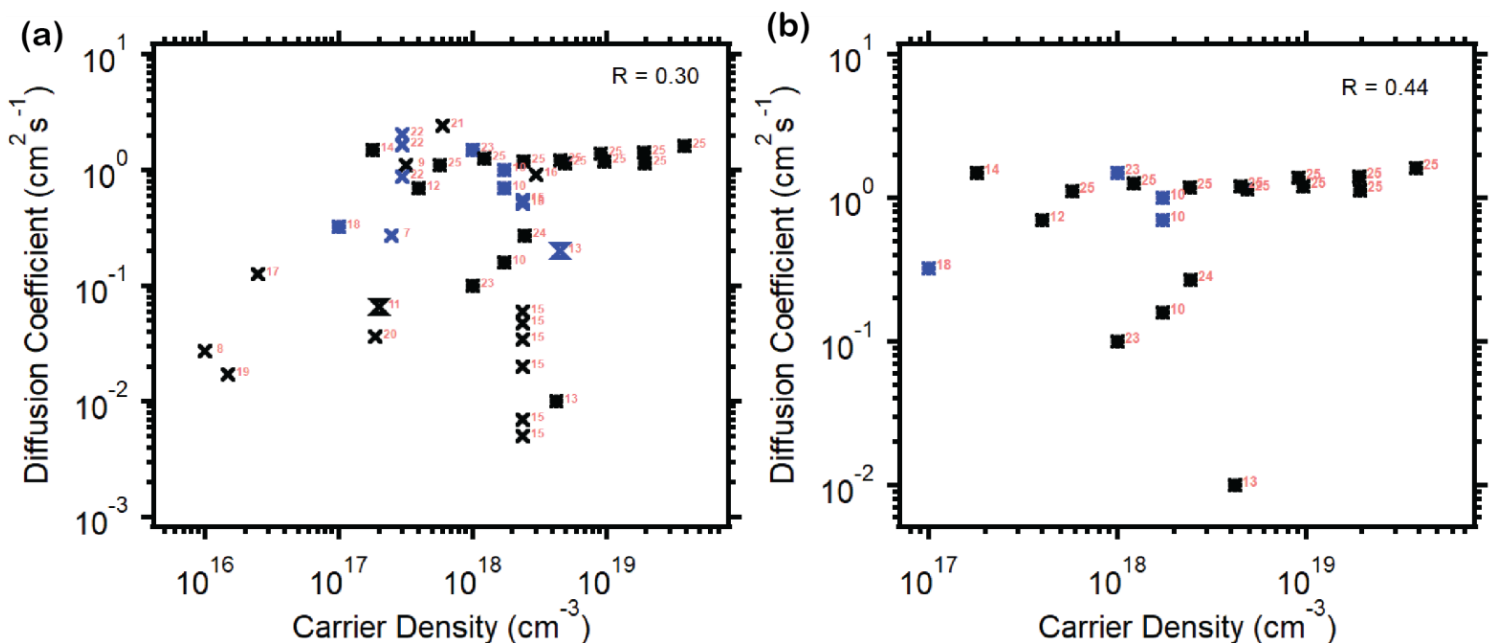

**Figure S5.** a) Literature survey of the reported diffusion coefficients as a function of initial carrier density ($N_0$) with a Pearson coefficient (R) of 0.30 (N = 43), indicating a weak positive correlation. Square data points indicate fits using the Mean Squared Displacement (MSD) Model, X-shaped data points indicate fits to a partial differential equation (PDE), hour-glasses-shaped data points indicate a fit using a combination of the MSD Model and a PDE. Black data points indicate $D$ values for polycrystalline films and blue data indicate $D$ values for single crystal samples. b) Data points for the MSD Model only with R = 0.44 (N = 20), indicating a weak positive correlation. The $N_0$ values range from $1\times10^{16}$ to $3.8\times10^{19}$ cm$^{-3}$ and the fitted diffusion coefficient values range from 0.005 to 2.4 cm$^2$ s$^{-1}$. Numbers to the right of the data points are the corresponding references.[7-25]

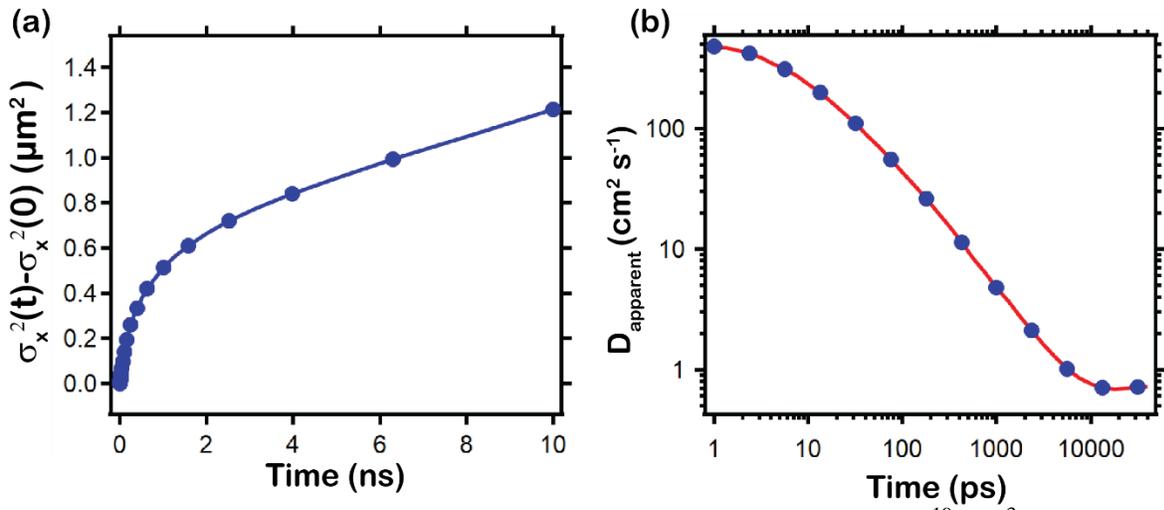

**Figure S6.** a) Mean squared displacement (MSD) plots for $N_{0,total} = 1 \times 10^{19}$ cm$^{-3}$ and initial spot size of 1000 nm. b) Apparent diffusion coefficient ($D_{app}$) as a function of time calculated by taking the instantaneous slope $\left(i.e. \frac{(MSD_{n+1} - MSD_n)}{(t_{n+1} - t_n)}\right)$ of the curve for $1 \times 10^{19}$ cm$^{-3}$ in a). Domain size = 15 x 15 $\mu$m; $k_1 = 5 \times 10^6$ s$^{-1}$; $k_2 = 2 \times 10^{-10}$ cm$^3$ s$^{-1}$; $k_3 = 1 \times 10^{-26}$ cm$^6$ s$^{-1}$; $D_{sim} = 0.7$ cm$^2$ s$^{-1}$.

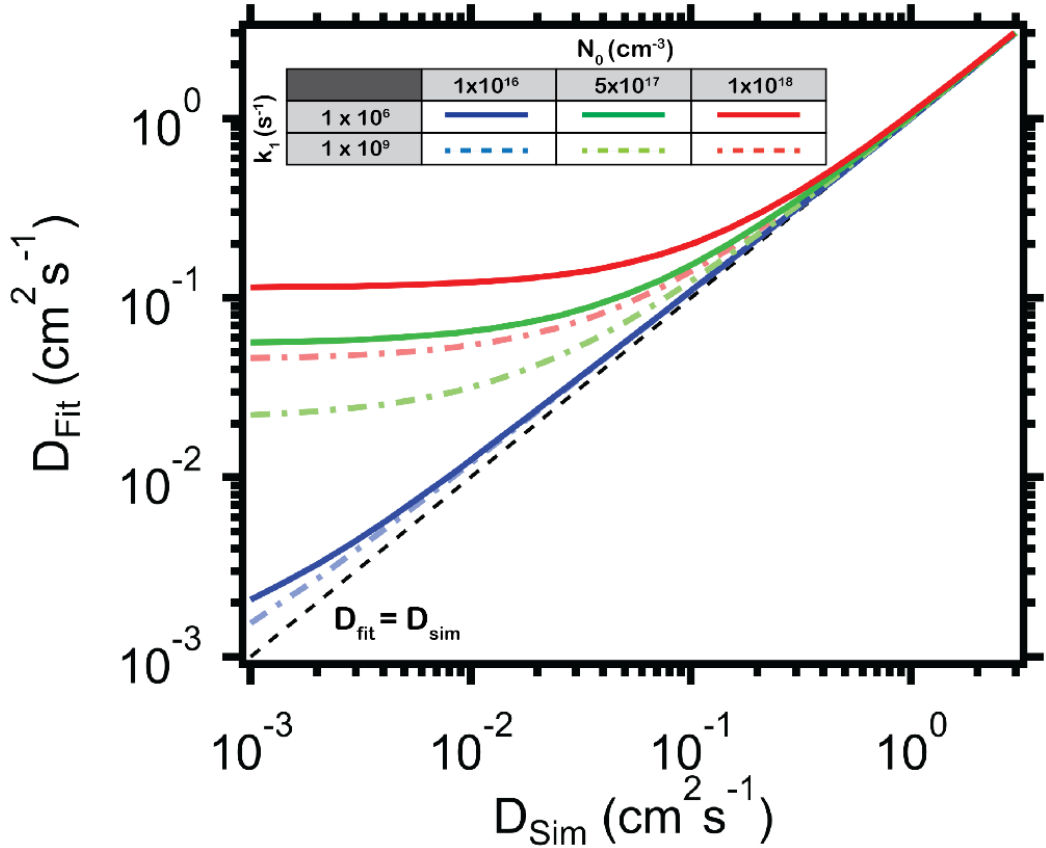

**Figure S7.** $D_{fit}$ using the Mean Squared Displacement Model as a function of $D_{sim}$ for $N_{0,total} = 1 \times 10^{16}$ cm$^{-3}$ (blue), $5 \times 10^{17}$ cm$^{-3}$ (green), $1 \times 10^{18}$ cm$^{-3}$ (red) when the non-radiative, first-order recombination constant ($k_1$) is set to $1 \times 10^6$ s$^{-1}$ (solid lines) versus $1 \times 10^9$ s$^{-1}$ (dashed lines). Fitting error is evident in both scenarios, indicating that the perovskite material quality does not significantly impact the key findings of this work.

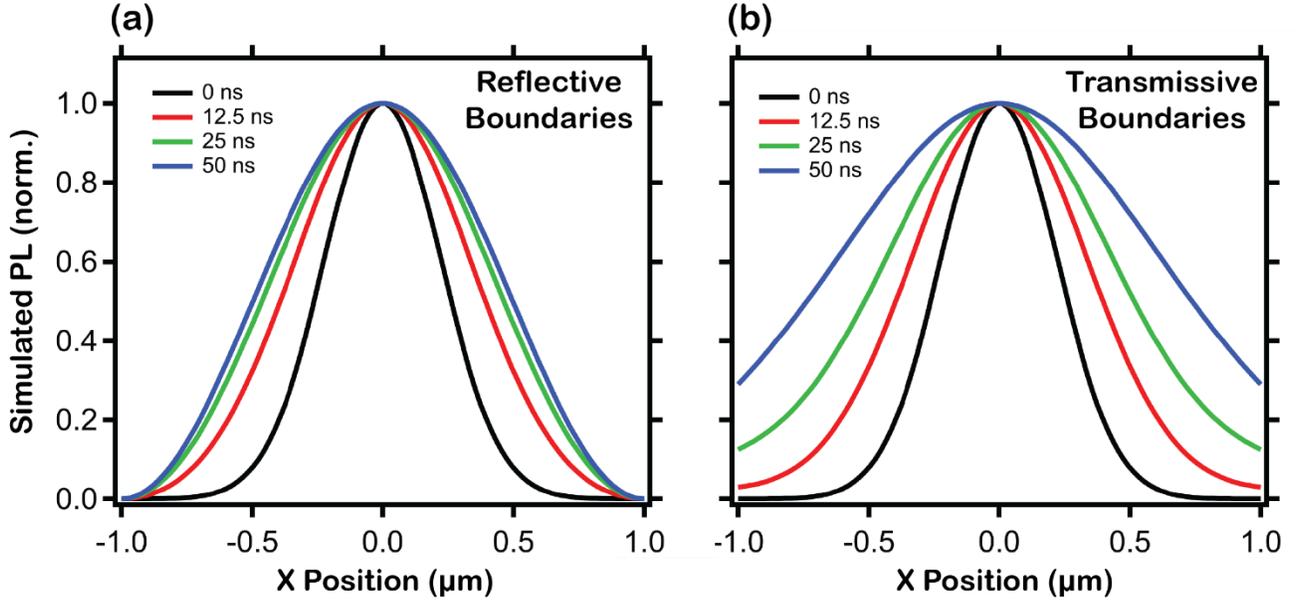

**Figure S8.** Time-dependent simulated photoluminescence (PL) line profiles along the *x*-direction at $t = 0$, 12.5, 25, and 50 ns for a) no flux (i.e. reflective, equation (S56)) compared to b) non-zero flux (i.e. transmissive and quenching, equation (S57)) boundary conditions. Domain size = 2 x 2 μm; $N_0 = 1\times10^{16}$ cm$^{-3}$; $k_1 = 1\times10^6$ s$^{-1}$; $k_2 = 2\times10^{-10}$ cm$^3$ s$^{-1}$; $k_3 = 1\times10^{-28}$ cm$^6$ s$^{-1}$; $D_{sim} = 0.05$ cm$^2$ s$^{-1}$ and S= 600 cm s$^{-1}$ for b). In (a) the sublinear change in the variance is due to the grain boundaries acting as solid walls and the nearly fixed carrier density profile decaying at a similar rate across the entire grain.

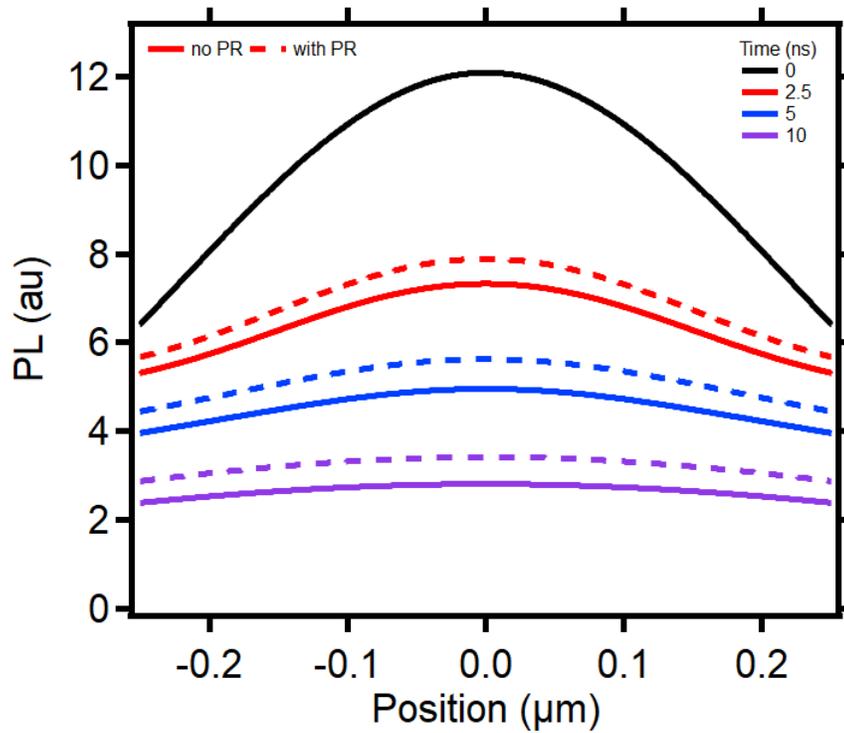

**Figure S9.** Absolute intensity temporal PL line profiles along the *x*-direction at 0, 2.5, 5, and 10 ns with (dashed lines) and without (solid lines) photon recycling from Figure 6 in the main article.

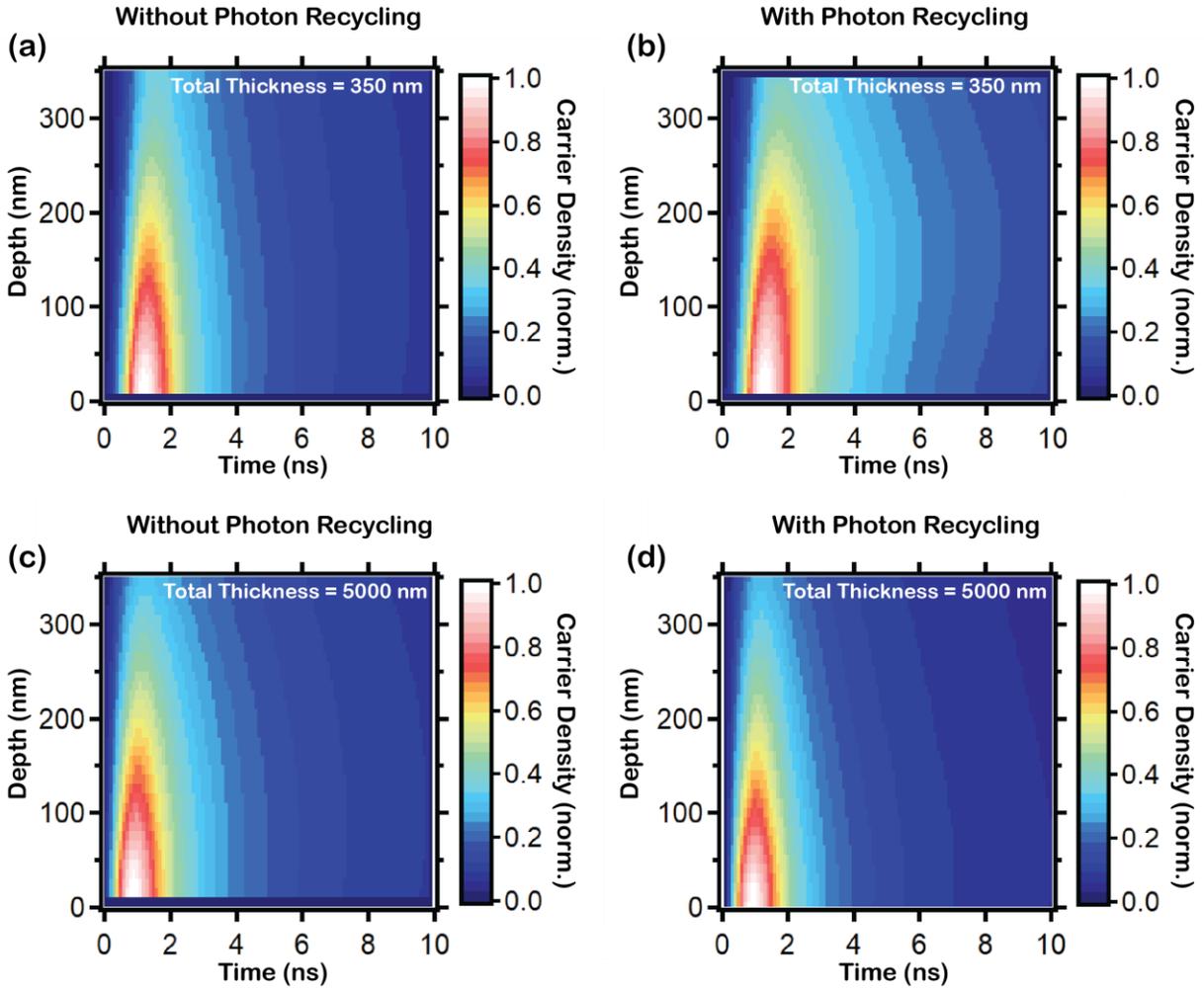

**Figure S10.** One-dimensional simulation of carrier density intensity as a function of time for a 350 nm thick film a) without and b) with photon recycling. This data reproduces the simulation by Ansari-Rad *et al.*.[4] Identical simulation c) without and d) with photon recycling for a 5000 nm thick film. Photon recycling only has an impact for the 350 nm film where photons rapidly reach the domain boundaries.

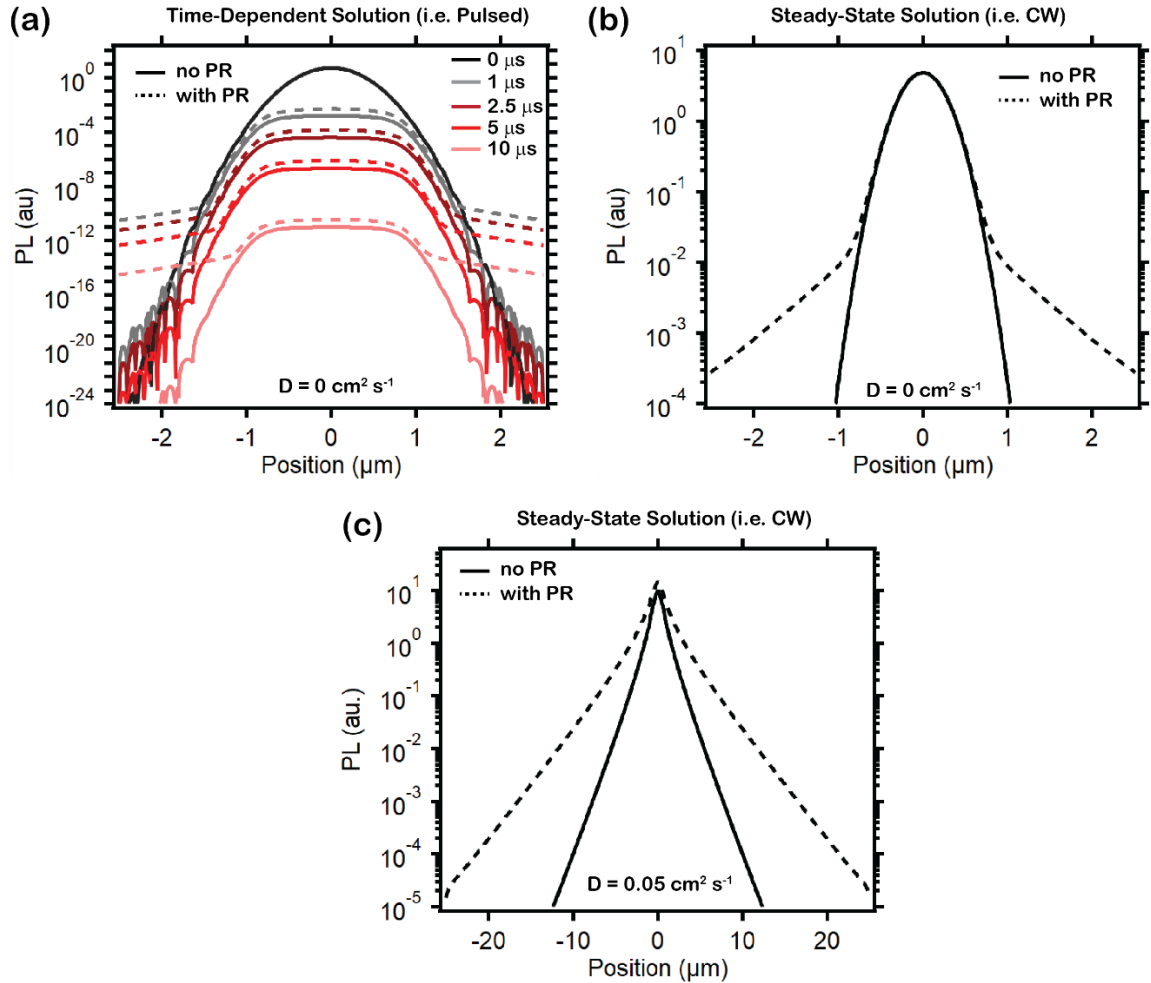

**Figure S11.** Simulated PL profiles, with photon recycling (dashed lines) and without photon recycling (solid lines) for a typical lead halide perovskite thin film with $k_1 = 1\times10^6$ s$^{-1}$; $k_2^{int} = 2\times10^{-10}$ cm$^3$ s$^{-1}$ with a) an initial Gaussian excitation pulse and no carrier diffusion; b) a constant generation source (i.e. CW) and no carrier diffusion; and c) CW with carrier diffusivity $D = 0.05$ cm$^2$ s$^{-1}$. Boundary conditions for photons are the same as those used by Ansari-Rad *et al.*.[4]

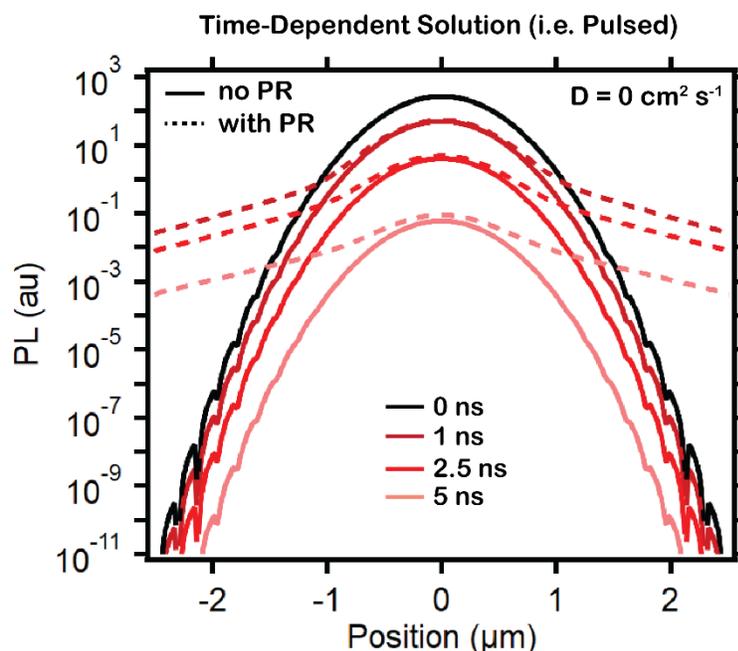

**Figure S12.** Simulated photoluminescence (PL) profiles, with photon recycling (dashed lines) and without photon recycling (solid lines) for an excitonic material with a radiative recombination rate constant of $k_R = 0.2 \times 10^9$ s$^{-1}$ and no carrier diffusion. The spectrally weighted absorption coefficient $\alpha_e = 1.4 \times 10^4$ cm$^{-1}$ and the average photon probability of escape from the film $P_{esc} = 7\%$ were kept the same as for a perovskite thin film.

**Supporting Information References**